\documentclass{JHEP3}
\usepackage{amsmath}
\usepackage{epsfig}

\newcommand{\be}{\begin{equation}}
\newcommand{\ee}{\end{equation}}
\newcommand{\bea}{\begin{eqnarray}}
\newcommand{\eea}{\end{eqnarray}}
\newcommand{\nn}{\nonumber}
\newcommand{\eps}{\epsilon}
\newcommand{\icalst}[1]{\mbox{${\cal S}^{-1}_{#1}$}}
\newcommand{\cals}{\mbox{${\cal S}$}}
\newcommand{\dbar}{\mbox{$\bar{\delta}$}}

\title
{Algebraic evaluation of rational polynomials \\ in one-loop amplitudes}
\author
{ T.~Binoth\\
School of Physics,
           The University of Edinburgh,\\
	   Edinburgh EH9 3JZ,
	   Scotland}
\author{J.~Ph.~Guillet\\
Laboratoire d'Annecy-le-Vieux de Physique Th\'eorique\\
LAPTH, B.P. 110, F-74941 Annecy-le-Vieux Cedex, France}
\author{G.~Heinrich\\
Institut f\"ur Theoretische Physik, 
Universit\"at Z\"urich,\\ Winterthurerstrasse 190, 
8057 Z\"urich, Switzerland}

\abstract{
One-loop amplitudes are to a large extent determined by 
their unitarity cuts in four dimensions. 
We show  that
the remaining rational terms can be obtained from the ultraviolet
behaviour of the amplitude, and determine universal form factors 
for these rational parts by applying 
reduction techniques to the Feynman diagrammatic representation
of the amplitude. The method is valid for massless and massive
internal particles.  We illustrate this method by evaluating the 
rational terms of the one-loop amplitudes for $gg\to H$, 
$\gamma\gamma\to \gamma\gamma$, $gg\to gg$,$\gamma\gamma \to ggg$ and
$\gamma\gamma\to \gamma\gamma\gamma\gamma$.
}

\preprint{ hep-ph/0609054\\
Edinburgh 2006/21\\
LAPTH-1160/06\\
ZU-TH-19/06}

\begin{document}

\section{Introduction}

The upcoming LHC experiment provides a strong 
phenomenological motivation~\cite{houches05} 
to develop formalisms which allow the successful evaluation of partonic
multi-leg processes at one loop. Recently, many different methods 
have been proposed to deal with this highly complex 
task.
Apart from algebraic reduction algorithms,
which for multi-leg processes lead to a proliferation of terms and 
motivate numerical~\cite{Soper:1999xk,Ferroglia:2002mz,Nagy:2003qn,Kurihara:2005ja,Anastasiou:2005cb,Czakon:2005rk} or 
semi-numerical~\cite{Binoth:2002xh,Grace,delAguila:2004nf,vanHameren:2005ed,Binoth:2005ff,Ellis:2005zh,Denner:2005nn,Binoth:2006rc,Binoth:2006mf} treatments at some stage, 
twistor-space inspired methods  
\cite{witten,BST,BCF7,BBST,altogether,BBCF,Bern:2005hh,Britto:2006sj} have stimulated a great deal of activity to complete 
the task -- initiated already more than ten years ago\,\cite{Bern:1994zx,BDDK,moreUnitarity} -- 
to evaluate one-loop amplitudes elegantly from their 
unitarity cuts~\cite{NeqFourSevenPoint,BCFII,NeqFourNMHV}. 
These algorithms are fully successful for supersymmetric amplitudes 
or special classes of helicity amplitudes, 
where the UV behaviour is tamed, but for general Standard Model amplitudes,  
it is difficult to obtain information on the so-called rational polynomials, 
which are induced 
by the ultraviolet behaviour of Feynman integrals. 
This was analysed in great detail by Bern, Dixon, Dunbar and
Kosower in~\cite{BDDK}, where  criteria for 4-dimensional cut constructibility 
of one-loop amplitudes were derived.
We will refer to these criteria, called 
 ``uniqueness result" in~\cite{BDDK}, as the ``BDDK-theorem" in the following.

The application of unitarity cuts to calculate 
the cut-constructible part of non-super\-symmetric amplitudes 
has seen a lot of progress recently,
and lead already to some remarkable results, as for example 
the cut constructible part
of the six gluon amplitude~\cite{Britto:2006sj}. 
This amplitude, including the rational parts,  
also has been evaluated numerically at a certain phase space point~\cite{Ellis:2006ss}. 

\medskip
  
Very recently, progress also has been made to determine the remaining
rational ambiguities within the unitarity-based method, 
using the so-called ``bootstrap approach"~\cite{Bern:2005hs,Bern:2005ji,Bern:2005cq,Berger:2006ci}.
These ideas have lead to the successful determination of previously unknown 
multi-leg QCD amplitudes~\cite{Forde:2005hh,Berger:2006vq,Berger:2006sh}. 

An alternative approach to the evaluation of the polynomial terms 
of helicity amplitudes in QCD 
is worked out in detail in~\cite{Xiao:2006vr,Su:2006vs,Xiao:2006vt}, 
where the missing rational parts of the so far unknown six gluon helicity 
amplitudes have been given. 
The authors have used reduction formulae in Feynman parameter space,
based on the work of~\cite{Bern:1993kr,Bern:1992em}. 
Their formalism is designed for gluon amplitudes and 
massless internal propagators.

\medskip

In this article, we show that the rational polynomials can be 
obtained in a general way, for massive as well as massless amplitudes  
and arbitrary external particles, 
in isolation from the cut-constructible parts, 
using Feynman diagrammatic reduction 
techniques. 
Our method is based on the tensor reduction formalism
presented in \cite{Binoth:2005ff}, where explicit representations for
tensor form factors have been derived. Projecting on the
ultraviolet sensitive part of these form factors, we
obtain the rational parts.  
The projection leads to a considerable 
simplification of the reduction cascade, 
which yields relatively compact expressions.
This defines a method which allows for the automated evaluation of 
the rational polynomials  of arbitrary one-loop 
amplitudes\footnote{Work on a similar subject has also 
appeared very recently in~\cite{Ossola:2006us}.}.

\medskip

In this sense, the two approaches -- construction of an amplitude 
by unitarity cuts and using Feynman diagrams 
together with tensor reduction --
can be considered as complementary: for the polynomial part, the 
 Feynman diagrammatic approach seems to be more straightforward, 
 while for the remaining parts of the amplitude the unitarity-based 
 method often leads  to  faster and more compact results.
 Such a combination of techniques already has been employed 
 in~\cite{Badger:2006us} to obtain the one-loop amplitude for a 
 Higgs boson plus four negative helicity gluons.
 Of course, for such a combined formalism to be universally 
 applicable, the generalisation of the 
unitarity based methods to massive internal propagators is required.

\medskip

The paper is organised as follows: First we give a definition
of rational or non-cut-constructible terms of general one-loop 
amplitudes in section 2. As an illustration, we calculate the 
polynomial terms of some 3, 4, 5- and 6-point amplitudes in section 3. 
Section 4 contains our conclusions. 
In the appendix we provide formulae which are useful for the 
extraction of rational polynomials of IR divergent amplitudes. 

\section{Rational terms of one-loop amplitudes}

In this section we propose a definition of the rational
part of a one-loop amplitude. 
To disentangle rational polynomials of infrared and ultraviolet origin, it is convenient 
to define the rational part with respect to the corresponding 
IR regulated amplitude. As an IR regulator, we use off-shell momenta 
for the external legs of our tensor integrals or masses for internal 
fermion lines. 
In this way it is guaranteed that all poles in $\epsilon$ will be of
ultraviolet nature and 
only terms which are related to ultraviolet divergences lead to 
rational polynomials in an amplitude.
It turns out that in the resulting expressions, the 
limits to the original kinematics are well defined.
Amplitudes which  contain dimensionally regulated infrared divergences 
could be used 
directly as a starting point as well, however in this case 
the two- and three-point functions have to be treated in a different way, 
as will be discussed in the appendix.

\subsection{Definition of rational parts of amplitudes}

Each $N$-point amplitude $\Gamma$ can be written as a linear combination
of tensor Feynman integrals in $n=4-2\epsilon$ dimensions. 
Schematically this can be denoted as 
\bea\label{defGamma1}
\Gamma &=& \sum C_{\mu_1\dots\mu_R}(n,\{s_{ij},m_k\})\; I_N^{n,\mu_1\dots\mu_R}(\{s_{ij},m_k\}) \;.
\eea
For our purposes, $\Gamma$ can be regarded as a complex valued function 
of the dimension $n$ and  
the kinematical invariants $s_{ij},m_k$.
The tensor integrals are defined in momentum space 
as\footnote{To make contact to eq.(2.1) of 
\cite{Binoth:2005ff}, we note that we have set $r_{a_1}=\ldots=r_{a_N}=0$ 
here for ease of notation.}
\begin{eqnarray}
I^{n,\,\mu_1\ldots\mu_R}_N(\{s_{ij},m_k\}) = 
\int \frac{d^n k}{i \, \pi^{n/2}}
\; \frac{k^{\mu_1}\,\dots  k^{\mu_R}}{
(q_1^2-m_1^2+i\delta)\dots (q_N^2-m_N^2+i\delta)}\;,
\label{eq0}
\end{eqnarray} 
where $q_j=k+r_j$ are the propagator momenta
and $r_j=p_1+\dots +p_j$ are sums of external momenta.  
In \cite{Binoth:2005ff} we have presented all the relevant formulae to perform 
a complete tensor reduction of a general  $N$-point tensor integral 
of rank $R\le N$. 
For $N\ge 6$ the reduction is purely algebraic, in the sense that these rank $R$ $N$-point 
functions decay into a linear combination of $(N-1)$-point functions of rank $R$--1. 
For $N\le 5$ all tensor form factors are evaluated in terms of an adequate
basis. We do not repeat these formulae here, 
but give the essential definitions to
keep the paper self-contained. 
Introducing Feynman parameters leads immediately to the following  
representation of the tensor integrals  \cite{davy,Binoth:1999sp,Giele:2004iy}
\bea
I^{n,\,\mu_1\ldots\mu_R}_N(\{s_{ij},m_k\}) 
& = & 
(-1)^R \sum_{m=0}^{[R/2]} \left( -\frac{1}{2} \right)^m
\sum_{j_1\cdots j_{r-2m}=1}^N \, 
\left[ 
 (g^{..})^{\otimes m}\,r_{j_1\cdot}^{\cdot} \cdots r_{j_R\cdot}^{\cdot}
\right]^{\{\mu_1\cdots\mu_R\}}
\nonumber\\
& & \;\;\;\;\;\;\;\;
\times I_N^{n+2m}(j_1 \ldots ,j_{R-2m}\,;\{s_{ij},m_k\})\;.
\label{davyd}
\eea 
The  objects $I_N^{n+2m}(j_1 \ldots ,j_{R-2m}\,;\{s_{ij},m_k\})$ 
are  scalar integrals in $D=n+2m\,\,(m=0,1,2,\ldots)$ dimensions 
with Feynman parameters in the numerator: 
\bea
&& I^D_N(j_1,\dots,j_R\,;\{s_{ij},m_k\}) = \nonumber\\&& \qquad\qquad
(-1)^N\Gamma(N-\frac{D}{2})\int \prod_{i=1}^N dz_i\,
\delta(1-\sum_{l=1}^N z_l)\,z_{j_1}\dots z_{j_R}\left(-\sum\limits_{k,l=1}^N \mathcal{S}_{kl} z_k z_l/2\right)^{D/2-N}
\nonumber\\
&& \mathcal{S}_{kl} = (r_l-r_k)^2 - m_l^2 - m_k^2\;.
\label{fofa}
\eea
This leads to
\bea\label{defGamma2}
\Gamma &=& \sum C(n,\{j_l\},\{s_{ij},m_k\}) I_N^{n+2m}(\{j_l\};\{s_{ij},m_k\}) \;,
\eea
where the sum runs over all different integrals making up the amplitude. 

\bigskip

Note that the coefficients $C(n,\dots)$ depend on the way the numerators of the
Feynman diagrams are evaluated, i.e. on the renormalisation scheme which defines 
the dimension of the Clifford algebra and the dimensionality of the internal and
external particles 
(for a more detailed discussion see \cite{Kunszt:1993sd,Catani:1996pk,Smith:2004ck}).
If chiral fermions are present or if one wants to define
helicity amplitudes for partonic processes, the presence of $\gamma_5$ makes it necessary to
distinguish 4-dimensional from $(n-4)$-dimensional contributions of the
Dirac algebra and $n$-dimensional vectors. 
We note that due to the  dimension splitting~\cite{Veltman:1988au} of the 
loop momentum $k=\hat k+\tilde k$, 
where $\hat k$ is 4-dimensional and $\tilde k$ is $(n-4)$-dimensional, 
 a few integrals with $\tilde k^2$-terms in the numerator 
have to be known, which     
can be mapped to higher dimensional integrals\,\cite{moreUnitarity,vanHameren:2005ed}
\bea \label{int_alpha}
\int \frac{d^nk}{i \pi^{n/2}} \frac{(\tilde k \cdot \tilde k)^\alpha}{(k^2-M^2)^N} &=& 
 (-1)^{\alpha}  \frac{\Gamma(\alpha -\epsilon)}{\Gamma(1 -\epsilon)}
 \frac{n-4}{2}\; 
I_N^{n+2\alpha}\nonumber\\ 
\int \frac{d^nk}{i \pi^{n/2}} \frac{(\tilde k \cdot \tilde k)^\alpha k^\mu k^\nu}{(k^2-M^2)^N} &=&
(-1)^{\alpha+1} \frac{\Gamma(\alpha -\epsilon)}{\Gamma(1 -\epsilon)}
g^{\mu\nu} \frac{n-4}{4} 
\frac{n+2\alpha}{n} \,I_{N}^{n+2+2\alpha} \;.
\eea
Therefore, these integrals contribute to the sum in eq.~(\ref{defGamma2}).

\medskip

The rational part of an amplitude in general 
stems from two different sources:
Firstly from a linear combination of $d=4$ terms
with the finite rational terms of Feynman parameter integrals, secondly from 
$(n-4)$-dimensional remnants of the Dirac algebra and the treatment of internal
particles, which combine with UV pole parts of Feynman parameter integrals. 
Thus we define the {\em rational} part $\mathcal{R}$ of an amplitude $\Gamma$ 
as\footnote{We assume that $\Gamma$ is either a genuinely UV finite or an UV 
renormalised amplitude. Note that counterterms can be expressed in 
terms of one- and two-point functions.} 
\bea
&&\mathcal{R}[\Gamma] 
= \sum C(4,\{j_l\},\{s_{ij},m_k\}) \mathcal{R}[I_N^{n+2m}(\{j_l \};\{s_{ij},m_k\})] \label{defR}\\
 && \qquad + (n-4) \, \sum \, C'(4,\{j_l\},\{s_{ij},m_k\})\, 
 \mathcal{P}[I_N^{n+2m}(\{j_l \};\{s_{ij},m_k\})]\nonumber\\
\mbox{with}&&\nonumber\\
&& C'(4,\{j_l\},\{s_{ij},m_k\}) =  \frac{d}{dn}C(n,\{j_l\},\{s_{ij},m_k\})\Big|_{n=4}\;\;.\nonumber
\eea
In eq.\,(\ref{defR}), $\mathcal{P}$ is the projector onto the  pole part of the 
argument, i.e. the $1/\epsilon$-term in the $\epsilon$-expansion. 
The action of  $\mathcal{R}$ on Feynman parameter integrals  is outlined 
 in more detail below. It is the coefficient $C'$ which governs  
 the renormalisation scheme dependence.

\vspace*{5mm}

In \cite{Binoth:2005ff} we have given explicitly all the necessary formulae to reduce
the scalar integrals with nontrivial numerators to $n$-dimensional
one-, two- and three-point functions, ($n$+2)-dimensional three-and four-point functions
and ($n$+4)-dimensional four-point functions. 
Schematically
\bea\label{tij}
I_N^{n+2m}(\{j_l\})&=&\sum  \, \beta_1\,I_1^n 
+ \sum  \,\beta_2(\{j_l\})\,I_2^n(1|j_1|j_1,j_2)\\
&&+\sum  \, \beta_{3a}(\{j_l\})\,I_3^n(1|j_1|j_1,j_2|j_1,j_2,j_3)
  +\sum  \, \beta_{3b}(\{j_l\})\, I_3^{n+2}(1|j_1)\nn\\
&& +\sum  \,  \beta_{4a}(\{j_l\})\,
I_4^{n+2}(1|j_1|j_1,j_2|j_1,j_2,j_3)+ 
\sum  \,  \beta_{4b}(j)\,I_4^{n+4}(1|j)
] \Big\}\;,\nn
\eea
where the arguments $(1|j_1|j_1,j_2|j_1,j_2,j_3)$ denote 
integrals  with up to three Feynman parameters in the numerator, and 
$I_N^n(1)$ are the genuine $N$-point  scalar integrals, which will be denoted simply by 
$I_N^n$ in the following. These integrals we call 
 the ``{\tt GOLEM} integral representation"\footnote{{\tt GOLEM} stands for ``General One-Loop Evaluator of Matrix elements" \cite{Binoth:2006mf}.}.
It is defined by the property that inverse Gram determinants can be completely avoided
by using such a representation. 
The coefficients 
$\beta_k(\{j_l\})$ are polynomial in the kinematical variables $\{s_{ij},m_k^2\}$,
they do not depend on the dimensionality $n$. 
The {\tt GOLEM} integral representation is a preferable starting point for a numerical
evaluation of one-loop amplitudes.
In algebraic approaches it is useful to reduce the {\tt GOLEM} integrals 
further to a smaller integral basis which allows for an easy isolation
of IR/UV divergencies.  A convenient choice is to express 
each {\tt GOLEM} integral by a linear combination of
the scalar integrals $I_1^n$, $I_2^n$, $I_3^n$, $I_4^{n+2}$.
All necessary formulae can be found in \cite{Binoth:2005ff}.
It is only in this further reduction step that an $n$-dependence 
enters into the coefficients:
\bea\label{ti}
I_N^{n+2m}(\{j_l\})&=&
\sum c_1(n)\,I_1^n +
  \sum c_2(n)\,I_2^n
+\sum c_{3}(n)\,I_3^n
 + \sum c_{4}(n)\,I_4^{n+2}
\;.
\eea
The summation is understood over the different kinematically
allowed 1,2,3- and 4-point functions, which 
are defined by all possible propagator pinches of the corresponding
$N$-point function on the left-hand side. 


\vspace*{5mm}

To completely define the operator $\mathcal{R}$ introduced above, 
we need to determine its action on a linear combination of  
1,2,3 and 4-point functions of the form given above. 
Using again the rule
\bea
\mathcal{R}[c(n) I_N] = c(4) \, \mathcal{R}[I_N] + 
(n-4)\, c'(4)\mathcal{P}[I_N]\;,
\label{defRc}
\eea
one sees that the rational part of an arbitrary  integral of type 
$I_N^{n+2m}(\{j_l\})$ 
is defined by the rational and the pole part of scalar integrals 
with trivial numerators.
As UV poles can only occur for 
$D/2+m\geq N$, 
all $D$=4 three-point and $D$=6 four-point functions are UV finite,
i.e. their pole parts are zero:  
\bea
\mathcal{P}[I_4^6]  = 0\;&,&\;\mathcal{P}[I_3^4]  = 0\;\;\;.   \label{ruleI3}
\eea
Further, the BDDK~\cite{BDDK} theorem tells us that 
all polynomial terms of such integrals, 
e.g. terms $\sim \pi^2/6$, are fully 
reconstructible by considering 4-dimensional cuts.
This suggests to {\em define} the rational parts
of these integrals to be zero: 
\bea
\mathcal{R}[I_3^4] = 0\;&,&\;\mathcal{R}[I_4^6] = 0\;\;\;. \label{ruleI4}
\eea
A detailed description of how these functions are uniquely reconstructible
from  the asymptotic logarithmic behaviour of an expression
can be found in \cite{Bern:1994zx,BDDK}.

\medskip

We define the pole- and rational parts of two-point functions
for $s\neq 0$ by considering the following massive representation
\bea
I_2^n(s,m_1^2,m_2^2) &=& \frac{\Gamma(1+\epsilon)}{\epsilon}
- \int\limits_{0}^{1} dx \log( -s x (1-x) + x m_1^2 + (1-x) m_2^2 )\nn
\eea
as
\bea
\mathcal{P}[I_2^n]  = \frac{1}{\epsilon}  \;&,&\; \mathcal{R}[I_2^n]  = 0\;.
\label{ruleI2}
\eea
Note that in the case $m_i=0$, another natural definition could be 
$\mathcal{R}[I_2^n]=2$, 
but these rational terms are directly related to the cut-constructible logarithmic
terms. 

\medskip

In the case $s=0$ the two-point function degenerates to combinations of 
one-point functions. The different cases are related to one-point functions
in the following way:
\bea\label{i2i1}
I_2^n(0,m_1^2,m_2^2) &=& \frac{1}{m_1^2-m_2^2} \Bigl( I_1^n(m_1^2) - I_1^n(m_2^2) \Bigr)\nonumber\\
I_2^n(0,0,m^2)       &=& \frac{2}{n-2} I_2^n(0,m^2,m^2) = \frac{1}{m^2} \; I_1^n( m^2 ) \; .
\eea
As up to $\mathcal{O}(\epsilon)$
\bea
I_1^n(m^2) &=& \frac{\Gamma(1+\epsilon)}{\epsilon\,(1-\epsilon)}\, m^2  - m^2\,\log( m^2 )\nn\;,
\eea
we define the
the pole and rational part of the one-point function to match
the non-logarithmic term, i.e.
\bea\label{ruleI1}
\mathcal{P}[I_1^n(m^2)]  = \frac{m^2}{\epsilon} \;&,&\; \mathcal{R}[I_1^n(m^2)]  = m^2\;.
\eea
Note that the rational parts of $I_2^n(0,0,m^2)$ and 
$I_2^n(0,m^2,m^2)$ turn out to be different using this definition, on the other hand
they still respect relation (\ref{i2i1}). 

\medskip

After having defined the pole and rational parts of the scalar integrals $I_4^{n+2}$, $I_3^n$, $I_2^n$
and $I_1^n$, the rational part of an amplitude, Eq.~(\ref{defR}), is fully determined. 
For a renormalised,  IR-finite amplitude,  
we are now in the position to define the {\em cut-constructible} part
of the amplitude $\Gamma$ indirectly by
\bea
\mathcal{C}[\Gamma] = ( 1  -\mathcal{R} )[\Gamma] \; .
\eea
In the general case one may use the definition
\bea
\mathcal{C}[\Gamma] = \lim_{\textrm{IR}} \, ( 1 - \mathcal{R} )[\Gamma_{\textrm{IR regulated}}] \; .
\eea
By an ``IR regulated" version of the amplitude we mean here a representation
where  internal propagator masses or virtualities of external particles which are 
zero in the original amplitude 
are treated as non-zero, i.e. $m_j\neq 0$ or 
$s_j=p_j\cdot p_j\neq 0$, in tensor form factors. This renders
the amplitude IR finite in the limit $n\to 4$ and the extraction of 
pole and rational parts is as in the IR finite case, if we demand $\lim_{\textrm{IR}}\,\Gamma_{\textrm{IR regulated}}=\Gamma$, where 
 $\lim_{\textrm{IR}}$ denotes the limits $m_j\to 0$ and/or $s_j\to 0$.
In doing so it is of course crucial to avoid the expansion in $(n-4)$,  
which does not commute with the limits $m_j\to 0, s_j\to 0$.
The IR-limits $m_j^2\to 0$ or $s_j\to 0$ do {\em not} commute in general 
with the extraction of pole and rational terms. 
An example will be discussed below in subsection \ref{subsec:4gluon}. 
In the appendix we provide explicit expressions
for pole and rational parts of IR divergent tensor form factors, 
which can be used 
to define rational terms differently, 
without using this IR regularisation procedure.
  
We have checked that in the case of massless internal particles 
our definition of cut-constructibility  
leads to identical form factor expressions as the ones 
presented in \cite{Xiao:2006vr} which were used to confirm known results
for five- and six-gluon amplitudes derived with the methods of \cite{BDDK}.  
We stress that it is not necessary to have an off-shell representation 
of the full amplitude; the IR-regulated representations 
of the needed tensor form factors are sufficient to define the rational
and pole parts we are looking for.  

For later use, we also define the following operator:
\be
{\cal U}[I]=(\mathcal{P}+\mathcal{R})[I]\;.
\label{defU}
\ee

Note that the unitarity based methods \cite{Bern:1993kr,Bern:1992em}
are not yet  generalised to the  case of massive loop integrals. 
Our definition can  easily be adapted to modified definitions of {\em cut-constructibility}
for massive one-loop amplitudes
by redefining rules (\ref{ruleI3})-(\ref{ruleI1}), once such a formalism is developed.

\medskip

For practical purposes, it is  more convenient to 
produce purely rational expressions for tensor integrals directly once and for all, 
instead of reducing to scalar integrals first and then 
apply the operators ${\cal P}$ and ${\cal R}$, because this avoids an 
explosion of terms at intermediate stages of the calculation. 
Therefore, we give a list of the pole- and
rational parts of the form factors for tensor integrals for $N=2,3$ and 4,  
which were needed in the applications below.  
As will be explained below, their knowledge is sufficient for 
the  determination of the rational parts of
any $N$-point amplitude, including amplitudes with massive internal particles.

\subsection{Rational parts of 2-point form factors}
Higher dimensional 2-point functions can be reduced to 
$(4-2\eps)$-dimensional ones by applying scalar integral reduction formulae \cite{Bern:1993kr,Bern:1992em,Binoth:1999sp,Binoth:2005ff,Xiao:2006vr}.
\bea
&& I_2^{n+2}(s,m_1^2,m_2^2)=\frac{1}{2s\, (n-1)} \nonumber\\
&& \qquad \times \; \Bigl[ (-s+m_1^2-m_2^2) I_1^n(m_2^2) 
+ (-s-m_1^2+m_2^2) I_1^n(m_1^2) +
\lambda(s,m_1^2,m_2^2)\, I_2^n(s,m_1^2,m_2^2)\Bigr]\; , \nonumber\\
&&\lambda(x,y,z)=x^2+y^2+z^2-2( xy +yz +xz )\;.
\label{i2np2}
\eea
Applying ${\cal U}$ to this formula and 
using rules (\ref{ruleI2}),(\ref{ruleI1}) yields, 
for the special case $m_1^2=m_2^2=m^2$,
\bea
{\cal U} [ I_2^{n+2}(s,m^2,m^2) ] &=&
\frac{s}{6} \Bigl( \frac{1}{\epsilon} + \frac{2}{3} \Bigr)
- m^2 \left( \frac{1}{\epsilon} + 1 \right) \;.
\eea
The rational parts of the 2-point form factors can
be read off directly from the formulae in appendix A of \cite{Binoth:2005ff}.
The form factors are defined by
\bea
I_2^{n,\,\mu}(s,m_1^2,m_2^2) 
& = &  r^{\mu} \, A^{2,1}\nn\\
I_2^{n,\,\mu_1 \mu_2}(s,m_1^2,m_2^2) 
& = & g^{\mu_1 \, \mu_2} \, B^{2,2} + 
r^{\mu_1} \, r^{\mu_2} \, A^{2,2}\nn
\eea
For equal internal masses one obtains 
\bea
{\cal U}[A^{2,1}(s,m^2,m^2)] &=& -\frac{1}{2\epsilon} \nonumber\\
{\cal U}[B^{2,2}(s,m^2,m^2)]&=& 
-\frac{s}{12} \left( \frac{1}{\epsilon} + \frac{2}{3} \right)
+\frac{m^2}{2} \left( \frac{1}{\epsilon} + 1 \right) \nonumber\\
{\cal U}[A^{2,2}(s,m^2,m^2)]&=& \frac{1}{3} \left( \frac{1}{\epsilon}+
\frac{1}{6} \right) \;.
\eea
Formulas for different internal masses are obtained analogously, 
but are not listed here as they are rather lengthy and because they are
not needed in the following.

\subsection{Rational parts of 3-point form factors}

Again, the rational parts can be obtained
by applying  ${\cal U}={\cal R}+{\cal P}$ as defined above 
to the reduction formulae in section 5.1 of \cite{Binoth:2005ff}.
In order to give compact formulae for the rational part of the three point form factors, 
we will introduce the following matrix:
\begin{equation}
H_{l_1 \, l_2} = \frac{b_{l_1} \, b_{l_2}}{B} - \icalst{l_1 \, l_2} \; , \; l_1,l_2 \in \{1,2,3\}\;,
\label{eqDEFHMAT}
\end{equation}
where $b_l=\sum_{k=1}^3 {\cal S}^{-1}_{lk}$ and $B=\sum_{k=1}^3 b_k$.
Note that internal propagator masses are present through the matrix ${\cal S}$
defined in eq. (\ref{fofa}).
Applying momentum conservation $r_3=p_1+p_2+p_3=0$ and defining $G_{lk}=2 \;r_l\cdot r_k$
for $l,k\in \{1,2\}$, it is easy to see that the third minor of $H$ is just $G^{-1}$: 
\bea
G^{-1} = -\frac{1}{\lambda(s_1,s_2,s_3)} 
\left( \begin{array}{cc} 2\, s_3 & s_2 - s_1 - s_3 \\
                         s_2 - s_1 - s_3  & 2\, s_1 \end{array} \right)
\eea
with $s_j = p_j\cdot p_j$ and $\lambda(s_1,s_2,s_3)$ as defined in eq.\,(\ref{i2np2}).
Then one can define the quantity 
\begin{eqnarray}
V_{l_1 \, l_2 \, l_3} & = & - \frac{1}{18} \, \Biggl[ 
H_{l_1 \, l_2} \, \left( \frac{1}{1+\delta_{l_2 \, l_3}} + \frac{1}{1+\delta_{l_1 \, l_3}} \right)  
 + 5 \, H_{l_1 \, l_2} \, \frac{b_{l_3}}{B} \; + 
1 \leftrightarrow 3 + 2 \leftrightarrow 3 
\Biggr]\,.
\label{eqDEFU3}
\end{eqnarray}
Here the ratios $b_j/B$, where
\bea
\frac{b_1}{B} = \frac{-s_3\,( s_1+s_2-s_3 ) + m_3^2\,( s_1-s_2-s_3 ) + 2\,s_3\,m_1^2 - m_2^2\,( s_1-s_2+s_3 )}{\lambda(s_1,s_2,s_3)}\nn\;,
\eea
and $b_2/B,b_3/B$ are obtained by cyclic permutations of indices, 
are well defined in the limit of massless internal propagators
and light-like on-shell kinematics, as they behave like $1/\lambda(s_1,s_2,s_3)$, 
which is well defined as long as at least one $s_j$ is non-zero.
Thus $H$ and $V$ are both well defined for all relevant kinematical cases, i.e. 
$m_j^2\to 0$ and/or  $s_j\to 0$.
Note however that these limits are {\it not} always equal to the finite parts of the 
corresponding integrals in the case where one or two of the $s_j$
are vanishing, as on-shell limits and $\eps$-expansion 
do not commute in general. 
This issue will be treated in detail in appendix \ref{sec:IR}.

The tensor form factors are defined by
\bea
&& I_3^{n,\,\mu_1}(r_1,r_2,r_3=0,m_1,m_2,m_3) = \sum\limits_{j_1=1}^{2} A^{3,1}_{j_1}\; r_{j_1}^{\mu_1} \nonumber\\
&& I_3^{n,\,\mu_1\mu_2} = (r_1,r_2,r_3=0,m_1,m_2,m_3) = \quad B^{3,2}\; g^{\mu_1\mu_2}  
\, +\sum\limits_{j_1,j_2=1}^{2} A^{3,2}_{j_1j_2}\; r_{j_1}^{\mu_1}r_{j_2}^{\mu_2} \nonumber\\
&&I_3^{n,\,\mu_1\mu_2\mu_3} = (r_1,r_2,r_3=0,m_1,m_2,m_3) = \nonumber\\
&& \qquad \sum\limits_{j_1=1}^{2} B^{3,3}_{j_1} \;( g^{\mu_1\mu_2} r_{j_1}^{\mu_3} + 2\; \mbox{perms.} ) 
\, +\sum\limits_{j_1,j_2,j_3=1}^{2} A^{3,3}_{j_1j_2j_3}\; r_{j_1}^{\mu_1}r_{j_2}^{\mu_2}r_{j_3}^{\mu_3}\;,
\eea
where we used momentum conservation to have $r_3=0$.
Using the equations in section 5.1  of ref.~\cite{Binoth:2005ff}, one gets the 
following pole- and rational parts of the form factors: 
\bea
{\cal U} [ A^{3,1}_{l} ]&=& 0 \nonumber\\
{\cal U} [ B^{3,2} ]&=& \frac{1}{4} 
\left(  \frac{1}{\epsilon} + 1 \right)\label{b32}\\
{\cal U} [ A^{3,2}_{l_1 \, l_2} ]&=& -\frac{1}{2}\,H_{l_1 \, l_2}  \nonumber\\
{\cal U} [ B^{3,3}_l ]&=& -\frac{1}{12} 
\left(  \frac{1}{\epsilon} + \frac{2}{3} + 
 \frac{b_l}{B} \right)\nonumber\\
{\cal U} [ A^{3,3}_{l_1 \, l_2 \, l_3} ]&=& -V_{l_1 \, l_2 \, l_3}\;.
\label{AB3}
\eea
The rational part of the rank one tensor integral is identically zero.
In the massless case these formulae are identical to the ones derived 
for massless internal particles in \cite{Xiao:2006vr}. 

The rational parts of the $(n+2)$-dimensional three-point functions are implicitly defined, e.g.
for the case of all masses equal one finds
\bea
{\cal U} [ I_3^{n+2}(s_1,s_2,s_3,m^2,m^2,m^2) ] &=&{\cal U} [-2\, B^{3,2}]
= -\frac{1}{2} \left(  \frac{1}{\epsilon} + 1 \right)\;.
\eea

\subsection{Rational parts of 4-point form factors}

From the preceding subsection and the way the form factors have been computed in 
ref.~\cite{Binoth:2005ff}, it is clear that only the rank 
3 and rank 4 form factors can have a non-zero rational part.
The form factors are defined by \cite{Binoth:2005ff}
\bea\label{4point_fofas}
&& I_4^{n,\,\mu_1\mu_2\mu_3}(r_1,r_2,r_3,r_4=0,m_1,m_2,m_3,m_4) = \nonumber\\
&& \qquad \sum\limits_{j_1=1}^{3} B^{4,3}_{j_1} \;( g^{\mu_1\mu_2} r_{j_1}^{\mu_3} + 2\; \mbox{perms.} ) 
\, +\sum\limits_{j_1,j_2,j_3=1}^{3} A^{4,3}_{j_1j_2j_3}\; r_{j_1}^{\mu_1}r_{j_2}^{\mu_2}r_{j_3}^{\mu_3} \\
&& I_4^{n,\,\mu_1\mu_2\mu_3\mu_4}(r_1,r_2,r_3,r_4=0,m_1,m_2,m_3,m_4) = 
C^{4,4} ( g^{\mu_1\mu_2} g^{\mu_3\mu_4} + 2\; \mbox{perms.} )\nonumber\\
&& \quad +\sum\limits_{j_1,j_2=1}^{3} B^{4,4}_{j_1j_2}\; ( g^{\mu_1\mu_2} r_{j_1}^{\mu_3} r_{j_2}^{\mu_4}+ 5\; \mbox{perms.} ) 
+\sum\limits_{j_1,j_2,j_3,j_4=1}^{3} A^{4,4}_{j_1j_2j_3j_4} \; r_{j_1}^{\mu_1}r_{j_2}^{\mu_2}r_{j_3}^{\mu_3}r_{j_4}^{\mu_4}\,.
\eea
For rank 3, we get

\bea
{\cal U}[ B^{4,3}_l ] &=& 0 \nn\\
{\cal U}[ A^{4,3}_{l_1 \, l_2 \, l_3} ] &=& - \frac{1}{6} \, 
\sum_{j \in S} \left[ H_{l_1 \, j} \, H^{\{j\}}_{l_2 \, l_3} \, \dbar_{j \, l_2} \, \dbar_{j \, l_3} + 
1 \leftrightarrow 2 + 1 \leftrightarrow 3 \right]
\eea
and for rank 4:
\bea
{\cal U}[ C^{4,4} ] & = & \frac{1}{24 \, \epsilon} + \frac{5}{72} \\
{\cal U}[ B^{4,4}_{l_1 l_2} ] & = & - \frac{1}{12 \, B}\,
\sum_{j \in S}  b_{j} \, H^{\{j\}}_{l_1 \, l_2} 
\, \dbar_{j \, l_1} \, \dbar_{j \, l_2} \\
{\cal U}[ A^{4,4}_{l_1 l_2 l_3 l_4} ] & = & 
f^{4,4}(l_1, l_2; l_3, l_4)+
f^{4,4}(l_1, l_3; l_2, l_4)+
f^{4,4}(l_1, l_4; l_3, l_2)\nn\\
&&+
f^{4,4}(l_2, l_3; l_1, l_4)+
f^{4,4}(l_2, l_4; l_3, l_1)+
f^{4,4}(l_3, l_4; l_1, l_2)\nn\\
&&+g^{4,4}(l_1; l_2, l_3, l_4)+
g^{4,4}(l_2; l_1, l_3, l_4)
\nn\\
&&+
g^{4,4}(l_3; l_2, l_1, l_4)+
g^{4,4}(l_4; l_2, l_3, l_1)\label{eqA44}\\
&&\nn\\
f^{4,4}(l_1, l_2; l_3, l_4)&=&
 \frac{1}{12 \, B} \, \sum_{j \in S}\, \dbar_{j \, l_3} \, \dbar_{j \, l_4}  
 \left[ H_{l_1 \, l_2} \, b_j + \frac{1}{2} \, b_{l_1} \, H_{j \, l_2} + 
 \frac{1}{2} \, b_{l_2} \, H_{l_1 \, j} \right]  \, H^{\{j\}}_{l_3 \, l_4} \nonumber \\
&&\nn\\
g^{4,4}(l_1; l_2, l_3, l_4)& =& 
- \frac{1}{4} \, \sum_{j \in S}  H_{l_1 \, j}  \, V^{\{j\}}_{l_2 \, l_3 \, l_4} \nn\\
&&\nn\\
\dbar_{j l} &=& 1- \delta_{j l} =  \left\{
\begin{array}{c}
1 \; \mbox{if} \; j \neq l \\
0 \; \mbox{if} \; j = l
\end{array}
\right. \nn\;.
\eea
In these formulae $S$ is the index set labelling the internal propagators
of the four-point function.
$H^{\{j\}}$ and $V^{\{j\}}$ are related to the  3-point kinematics obtained
when omitting propagator $j$ from this set\,\cite{Binoth:2005ff}.  
For completeness we  also list
\bea
{\cal U}[ I_4^{n} ] = 0 \;,\quad
{\cal U}[ I_4^{n+2} ] = 0 \;,\quad
{\cal U}[ I_4^{n+4} ] = \frac{1}{6\eps} + \frac{5}{18} \;\;.
\eea


\subsection{Rational parts of 5-point form factors}
With the results given above and the explicit representations
of the 5-point form factors given in \cite{Binoth:2005ff},  
it is manifest that up to rank
3, all 5-point functions have vanishing rational terms:
\bea
{\cal U}[ I_5^n ] = 
{\cal U}[ I_5^{n,\mu_1} ] =
{\cal U}[ I_5^{n,\mu_1\mu_2} ] = 
{\cal U}[ I_5^{n,\mu_1\mu_2\mu_3} ] = 0\;.
\eea
The rational terms of the form factors for the  rank 4 and rank 5
tensor integrals can directly be obtained from  the explicit
formulae given in section 6 of \cite{Binoth:2005ff},
which manifestly show that no integrals apart from the ones 
given in the previous subsections  appear in the reduction. 
For $N\geq 6$, the reduction is purely algebraic anyway, i.e. 
involves only kinematic matrices until five-point functions are reached.
Therefore, the formulae given in the 
previous subsections are sufficient to calculate the rational parts of 
{\it arbitrary} $N$-point amplitudes. 

As was shown in \cite{Binoth:2005ff},  $N$-point functions up to 
rank $N$--2 are 
algebraically reducible to rank 3 five-point functions,  
therefore all the corresponding
rational terms are zero. We have thus re-derived, within our
formalism, a well known result of \cite{BDDK}: all $N$-point amplitudes 
which contain, in a convenient gauge, at most rank $N$--2 tensor integrals
are cut-constructible.

\section{Applications}

In the following, five examples will be presented to illustrate 
our approach. The first is Higgs production by gluon fusion, 
the second scattering of light-by-light. Then we consider the
4-gluon amplitude which is IR and UV divergent.
Finally we discuss
a pentagon and a hexagon amplitude, 
$ggg\gamma\gamma\to 0$ and  $\gamma\gamma\gamma\gamma\gamma\gamma\to 0$.

\subsection{Example 1: Higgs production by gluon fusion}

Higgs production by gluon fusion is mediated by massive quark loops.
In the Standard Model, the top quark provides the leading contribution.
Up to a trivial colour structure $\sim \delta_{ab}$ the amplitude is given by 
\bea
\mathcal{M} &=& - \frac{m_t}{v}\frac{g_s^2}{(4\pi)^{n/2}} \int \frac{d^nk}{i \,\pi^{n/2}}
\frac{ \textrm{tr}( \varepsilon_1\!\!\!\!\!/\; (q_1\!\!\!\!\!/+m_t) (q_2\!\!\!\!\!/+m_t) \varepsilon_2\!\!\!\!\!/ \;(k\!\!\!/+m_t))
}{(q_1^2-m_t^2)(q_2^2-m_t^2)(k^2-m_t^2)} 
\eea  
Here $q_j=k+r_j$ with $r_1=-p_1$ and  $r_2=p_2$, where $p_j$ are the light-like 
momenta of the gluons with polarisation vectors $\varepsilon_j$.
Apart from $m_t^2$ the only non-vanishing Lorentz invariant variable is $2 p_1\cdot p_2=s=m_H^2$.
Working out the trace leads to 
\bea
\mathcal{M} &\sim& I_3^n( r_1,r_2,0,m_t^2,m_t^2,m_t^2 )\, ( -\varepsilon_1\cdot\varepsilon_2 
 + 2\,\varepsilon_1\cdot p_2\, \varepsilon_2\cdot p_1  + 2\, m_t^2 \, \varepsilon_1\cdot\varepsilon_2 )
 \nonumber \\ && + I_3^{n\,\mu\nu}( r_1,r_2,0,m_t^2,m_t^2,m_t^2 ) \, (  8 \,\varepsilon_{1\,\mu}\varepsilon_{2\,\nu} 
 - 2 \,\varepsilon_1\cdot\varepsilon_2\, g_{\mu\nu} )
\eea
The scalar integral $I_3^n$ does not contribute to the rational part 
of the amplitude, while the rank 2 tensor integral does. 
The pole and rational parts of the  form factors for $I_3^{n\,\mu\nu}$ defined above
turn out to be
\bea
&& {\cal U}[B^{3,2}] =  \frac{1+\epsilon}{4\,\epsilon} \nonumber \\
&& {\cal U}[A^{3,2}_{11}] = -{\cal U}[A^{3,2}_{12}] =
{\cal U}[A^{3,2}_{22}] =-\frac{1}{2 \, s}  \;.
\eea   
The decomposition of the Feynman diagram in terms of partly UV divergent tensor integrals
made it necessary to work in dimensional regularisation, although the amplitude is finite. 
The rational parts of the
amplitude are a result of products of UV $1/\epsilon$ poles and order  $\epsilon$ remnants from the
$n$-dimensional gamma algebra.
Note that the rational part of the tensor functions is only a small part of the
whole tensor integral.
Applying the rules of  section 2, the rational part is found to be
\bea
{\cal R}[\mathcal{M}] &=& \frac{\alpha_s}{\pi}\, \frac{m_t^2}{v} \; \frac{\textrm{tr}(\mathcal{F}_1\mathcal{F}_2)}{s}\;. 
\eea
Here  ${\cal F}_j^{\mu\nu}=p_j^\mu\varepsilon_j^\nu-p_j^\nu\varepsilon_j^\mu$ is the abelian part
of the gluon field strength tensor.

\subsection{Example 2: Scattering of light-by-light}

In QED, scattering of light-by-light is mediated by a closed electron loop. 
It is well known that the six box topologies making up the amplitude 
are related, such that it 
is sufficient to evaluate only one diagram
with a given ordering of the external photons. The others can be obtained by
all non-cyclic permutations of the photon momentum and polarisation vectors. 
\bea
\mathcal{M} &=& \frac{e^4}{(4\pi)^{n/2}} \sum\limits_{\sigma \in S_4/Z_4} \mathcal{G}(\sigma_1,\sigma_2,\sigma_3,\sigma_4) 
\eea 
In addition, due to charge invariance, one has $\mathcal{G}(1,2,3,4)=\mathcal{G}(1,4,3,2)$, 
$\mathcal{G}(1,3,4,2)=\mathcal{G}(1,2,4,3)$ and $\mathcal{G}(1,4,2,3)=\mathcal{G}(1,3,2,4)$. 
We  evaluate now the rational polynomial of $\mathcal{M}$ using 
the polynomial tensor coefficients.
The evaluation of the diagram
\bea\label{g1234}
\mathcal{G}(1,2,3,4)= -\int \frac{d^nk}{i \,\pi^{n/2}} 
\frac{ \textrm{tr}( \varepsilon_1\!\!\!\!\!/ \;(q_1\!\!\!\!\!/+m_e)\varepsilon_2\!\!\!\!\!/ \;(q_2\!\!\!\!\!/+m_e) \varepsilon_3\!\!\!\!\!/ 
\;(q_3\!\!\!\!\!/+m_e) \varepsilon_4\!\!\!\!\!/ \;(k\!\!\!/+m_e))}{
(q_1^2-m_e^2)(q_2^2-m_e^2)(q_3^2-m_e^2)(k^2-m_e^2)} 
\eea
involves four-point tensor integrals up to rank four.
These are in general complicated
functions, but the rational polynomials are actually very simple. Note that we are only interested
in the massless case $m_e\to 0$. We only keep the electron mass 
as an infrared cutoff for the moment.
Otherwise the massless on-shell 2-point functions, which are zero in dimensional
regularisation, would spoil a clear separation of IR and UV problems.
We give now the complete list of the polynomial part of the four-point tensor coefficients.
For rank zero, one and two, no rational terms are present. 
The rank 3 and rank 4 tensor coefficients are defined in eq. (\ref{4point_fofas}) above.
The rational polynomials of the tensor coefficients depend on the external kinematics. In our case, where
$p_j\cdot p_j=0$ for $j=1,2,3,4$, we find in the limit $m_e\to 0$ 
\bea
&& {\cal U}[ B^{4,3}_{1} ] =  {\cal U}[ B^{4,3}_2 ] = 
{\cal U}[ B^{4,3}_3 ] = 0 \nonumber\\
&& {\cal U}[ A^{4,3}_{111}] = {\cal U}[ A^{4,3}_{333}] = \frac{u-t}{2\,s\,t\,u}\nonumber\\
&& {\cal U}[ A^{4,3}_{112}] = {\cal U}[ A^{4,3}_{233}] =\frac{1}{2\,s\,u}\nonumber\\
&& {\cal U}[ A^{4,3}_{113}] ={\cal U}[ A^{4,3}_{122}] = {\cal U}[ A^{4,3}_{133}] = 
{\cal U}[ A^{4,3}_{223}]
= -{\cal U}[ A^{4,3}_{123}] = \frac{1}{2\,t\,u}\nonumber\\
&& {\cal U}[ A^{4,3}_{222}] = \frac{u-s}{2\,s\,t\,u}\;,
\eea
the remaining ones are defined by symmetry under exchange of the lower indices.
The non cut-constructible parts of the rank four form factors are 
\bea
&& {\cal U}[ C^{4,4} ] =  \frac{1}{24} \frac{1}{\epsilon} + \frac{5}{72}  \nonumber\\
&& {\cal U}[ B^{4,4}_{11} ] = {\cal U}[ B^{4,4}_{13} ] = {\cal U}[ B^{4,4}_{22} ] 
 = {\cal U}[ B^{4,4}_{33} ] = -{\cal U}[ B^{4,4}_{12} ] = -{\cal U}[ B^{4,4}_{23} ] = -\frac{1}{12\,u}\nonumber\\
&&{\cal U}[ A^{4,4}_{1111} ] = {\cal U}[ A^{4,4}_{3333} ] = \frac{1}{s\,t} - \frac{1}{s\,u} + \frac{1}{2\,u^2}\nonumber\\
&&{\cal U}[ A^{4,4}_{1112} ] = {\cal U}[ A^{4,4}_{2333} ] =\frac{1}{2\,s\,u} - \frac{1}{2\,u^2}\nonumber\\ 
&&{\cal U}[ A^{4,4}_{1113} ] = {\cal U}[ A^{4,4}_{1333} ] = - \frac{1}{2\,s\,t} - \frac{1}{2\,s\,u} + \frac{1}{2\,u^2}\nonumber\\
&&{\cal U}[ A^{4,4}_{1122} ] ={\cal U}[ A^{4,4}_{2233} ] = - \frac{1}{6\,s\,t} + \frac{1}{2\,u^2}\nonumber\\
&&{\cal U}[ A^{4,4}_{1123} ] ={\cal U}[ A^{4,4}_{1233} ] =\frac{1}{6\,s\,t} + \frac{1}{6\,s\,u} - \frac{1}{2\,u^2}\nonumber\\
&&{\cal U}[ A^{4,4}_{1133} ] =- \frac{1}{3\,s\,t} - \frac{1}{3\,s\,u} + \frac{1}{2\,u^2}\nonumber\\
&&{\cal U}[ A^{4,4}_{1222} ] ={\cal U}[ A^{4,4}_{2223} ] = - \frac{1}{2\,s\,t} - \frac{1}{2\,s\,u} - \frac{1}{2\,u^2}\nonumber\\
&&{\cal U}[ A^{4,4}_{1223} ] =\frac{1}{6\,s\,t} + \frac{1}{6\,s\,u} + \frac{1}{2\,u^2}\;,
\eea
the remaining ones are defined by symmetry. 
Evaluation of the rational part of eq.~(\ref{g1234}) is now straightforward.
Using these form factors we find for the rational part of the sum of all graphs
the manifestly gauge invariant result
\bea
&& {\cal R}[\sum\limits_{\sigma \in S_4/Z_4} \mathcal{G}(\sigma_1,\sigma_2,\sigma_3,\sigma_4)] \nn\\
&& \hspace{1cm} = \; \frac{8}{3} \frac{\textrm{tr}({\cal F}_1{\cal F}_2)}{s}  \frac{\textrm{tr}({\cal F}_3{\cal F}_4) }{s}
 + \frac{8}{3} \frac{\textrm{tr}({\cal F}_1{\cal F}_3)}{u}  \frac{\textrm{tr}({\cal F}_2{\cal F}_4) }{u}
 + \frac{8}{3} \frac{\textrm{tr}({\cal F}_1{\cal F}_4)}{t}  \frac{\textrm{tr}({\cal F}_2{\cal F}_3) }{t}\nn\\
&& \hspace{1cm}+  \frac{64}{3} \frac{\textrm{tr}({\cal F}_1{\cal F}_2)}{s} \frac{p_4\cdot{\cal F}_3\cdot p_1 \;\; p_3\cdot{\cal F}_4\cdot p_1}{s\, t\, u}
 -  \frac{64}{3} \frac{\textrm{tr}({\cal F}_1{\cal F}_3)}{u} \frac{p_1\cdot{\cal F}_2\cdot p_3 \;\;  p_3\cdot{\cal F}_4\cdot p_1}{s\, t\, u}
 \nn\\&&\hspace{1cm}
 +  \frac{64}{3} \frac{\textrm{tr}({\cal F}_1{\cal F}_4)}{t} \frac{p_1\cdot{\cal F}_2\cdot p_3 \;\;  p_4\cdot{\cal F}_3\cdot p_1}{s\, t\, u}
 +  \frac{64}{3} \frac{\textrm{tr}({\cal F}_2{\cal F}_3)}{t} \frac{p_2\cdot{\cal F}_1\cdot p_3 \;\;  p_3\cdot{\cal F}_4\cdot p_1}{s\, t\, u}
 \nn\\&&\hspace{1cm}
 -  \frac{64}{3} \frac{\textrm{tr}({\cal F}_2{\cal F}_4)}{u} \frac{p_2\cdot{\cal F}_1\cdot p_3 \;\;  p_4\cdot{\cal F}_3\cdot p_1}{s\, t\, u}
 +  \frac{64}{3} \frac{\textrm{tr}({\cal F}_3{\cal F}_4)}{s} \frac{p_2\cdot{\cal F}_1\cdot p_3 \;\;  p_1\cdot{\cal F}_2\cdot p_3}{s\, t\, u}
\nn\\&&\hspace{1cm}
  + 1024\; \frac{p_2\cdot{\cal F}_1\cdot p_3 \;\; p_1\cdot{\cal F}_2\cdot p_3}{s\, t\, u}\frac{p_4\cdot{\cal F}_3\cdot p_1 \;\; p_3\cdot{\cal F}_4\cdot p_1}{s\, t\, u}\;,
\eea
where ${\cal F}_j^{\mu\nu}=p_j^\mu\varepsilon_j^\nu-p_j^\nu\varepsilon_j^\mu$ is the electromagnetic field strength tensor.
On the amplitude level the UV pole cancels and thus ${\cal R} = {\cal U}$. 
The result simplifies further if one specialises to helicity 
amplitudes using spinor helicity methods \cite{xuetal}.
Due to parity invariance and Bose symmetry only the helicities $++++$, $+++-$
and $++--$ have to be considered. 

The rational polynomials of the three helicity amplitudes 
are then given by 
\bea
{\cal R}[\mathcal{M}^{++++}] = \quad 8 \, \alpha^2\: \varepsilon_1^{+}\cdot\varepsilon_2^{+} \,\varepsilon_3^{+}\cdot\varepsilon_4^{+} \hspace{1.5cm}
&=& \quad 8\, \alpha^2\: \frac{[12][34]}{\langle 12 \rangle \langle 34\rangle}\nonumber\\
{\cal R}[\mathcal{M}^{+++-}] = \quad 8 \, \alpha^2\: \varepsilon_1^{+}\cdot\varepsilon_2^{+}\,
           \varepsilon_3^{+}\cdot p_1\varepsilon_4^{-}\cdot p_1 \,\frac{2\, s}{t\,u} 
&=& -8\, \alpha^2 \:\frac{[12]\langle 14\rangle [13]}{\langle 12\rangle [14]\langle 13\rangle} \nonumber\\
{\cal R}[\mathcal{M}^{++--}] = -8 \, \alpha^2\: \varepsilon_1^{+}\cdot\varepsilon_2^{+}\,\varepsilon_3^{-}\cdot\varepsilon_4^{-} \hspace{1.5cm}
  &=&-8 \, \alpha^2 \:\frac{[12]\langle 34\rangle}{\langle 12 \rangle [34]}\;.
\eea
We note that up to a phase the rational parts of the different amplitudes are the same.
For the first two helicity configurations, the rational terms are the full result. For completeness we also 
quote the full result of the $++--$ case\,\cite{Gastmans:1990xh,Binoth:2002xg}:
\bea
\mathcal{M}^{++--} = -8 \, \alpha^2\: \left( 1
+\frac{t-u}{s}\log\left(\frac{t}{u}\right) 
+ \frac{1}{2}\frac{u^2+t^2}{s^2} \Bigl[\log^2  \left(\frac{t}{u}\right) +  \pi^2 \Bigr] \right) \:\frac{[12]\langle 34\rangle}{\langle 12 \rangle [34]}\;.
\eea

\subsection{Example 3: Gluon-gluon scattering}
\label{subsec:4gluon}
To apply our formalism to an IR divergent amplitude we
have chosen the 4-gluon amplitude as an example.
In the following we consider the  helicity amplitudes
$++++$ and $++--$. Using spinor helicity methods one can replace
the polarisation vectors by polynomial terms in the external momenta times a
global phase [$\textrm{tr}^\pm(\dots)=\textrm{tr}(\dots)/2 \pm \textrm{tr}(\gamma_5\dots)/2$]:
\bea
\varepsilon_{1}^{+\;\mu_1} \varepsilon_{2}^{+\;\mu_2}
\varepsilon_{3}^{+\;\mu_3} \varepsilon_{4}^{+\;\mu_4}&=& \frac{[21]}{\langle 12\rangle}\frac{[43]}{\langle 34\rangle} \frac{1}{2\,s^2} 
\textrm{tr}^-(\, k_1\!\!\!\!\!/ \,\gamma^{\mu_2} \,k_2\!\!\!\!\!/ \,\gamma^{\mu_1})\; \textrm{tr}^-(\,k_3\!\!\!\!\!/ \,\gamma^{\mu_4} \,k_4\!\!\!\!\!/\, \gamma^{\mu_3})\nonumber\\
\varepsilon_{1}^{+\;\mu_1} \varepsilon_{2}^{+\;\mu_2}
\varepsilon_{3}^{-\;\mu_3} \varepsilon_{4}^{-\;\mu_4}&=& \frac{[21]}{\langle 12\rangle}\frac{\langle 34\rangle}{[43]} \frac{1}{2\,s^2} 
\textrm{tr}^-(\, k_1\!\!\!\!\!/ \,\gamma^{\mu_2} \,k_2\!\!\!\!\!/ \,\gamma^{\mu_1})\; \textrm{tr}^+(\,k_3\!\!\!\!\!/ \,\gamma^{\mu_4} \,k_4\!\!\!\!\!/\, \gamma^{\mu_3})\;.
\eea
As we are using the 't\,Hooft-Veltman scheme for the computation, these expressions implicitly 
project onto the 4-dimensional part of $n$-dimensional objects in loop diagrams.  
At tree level one finds 
\bea
\mathcal{A}^{++++}_{\textrm{LO}} &=& \mathcal{A}^{+++-}_{\textrm{LO}} = 0 \nonumber\\ 
\mathcal{A}^{++--}_{\textrm{LO}} &=& -16\pi\,\alpha_s \frac{1}{N_C} \frac{[21]\langle 34\rangle}{\langle 12\rangle[43]}\hat{\mathcal{A}}^{++--}_{\textrm{LO}} \nonumber\\
\hat{\mathcal{A}}^{++--}_{\textrm{LO}} &=& 
    + \mathcal{T}_{\textrm{ad}}^{1234} \, \frac{s}{t}
    + \mathcal{T}_{\textrm{ad}}^{1243} \, \frac{s}{u} 
    + \mathcal{T}_{\textrm{ad}}^{1324} \, \frac{s^2}{tu}\;,
\eea
with the colour objects defined through a trace of colour matrices 
in the adjoint representation $T^c_{ab}=-i f^{cab}$
  \bea
\mathcal{T}_{\textrm{ad}}^{1234} =  \textrm{tr}(T^{c_1}T^{c_2}T^{c_3}T^{c_4})  \; .
\eea 
We have computed the next-to-leading order contributions in two ways to outline the extraction
of the rational terms for the IR divergent and IR regulated case. As the $++++$ case is
IR and UV finite, both evaluations give the same result for this helicity configuration:
\bea
\mathcal{A}^{++++}_{\textrm{NLO}} &=& -4\,\alpha_s^2 \frac{[21][43]}{\langle 12\rangle\langle 34\rangle}\hat{\mathcal{A}}^{++++}_{\textrm{NLO}}\nonumber
\nonumber\\
\hat{\mathcal{A}}^{++++}_{\textrm{NLO}} 
&=& \frac{1}{3} (  \mathcal{T}_{\textrm{ad}}^{1234} +  \mathcal{T}_{\textrm{ad}}^{1324}+  \mathcal{T}_{\textrm{ad}}^{1342} )\;.
\eea
The amplitude is defined by rational parts only.
In the $++--$ case we again extract an overall factor:
\bea
\mathcal{A}^{++--}_{\textrm{NLO}} &=& -4\,\alpha_s^2 \frac{[21]\langle 34\rangle}{\langle 12\rangle[43]}\hat{\mathcal{A}}^{++--}_{\textrm{NLO}}
\nonumber\\
\hat{\mathcal{A}}^{++--}_{\textrm{NLO}} &=& 
 + \frac{(s^2+t^2+u^2)^2}{2\,stu} \,\mathcal{T}_{\textrm{ad}}^{1324}  \, I_4^6(u,t)
 + 2\frac{su}{t} \, \mathcal{T}_{\textrm{ad}}^{1234} \, I_4^6(s,t) 
 + 2\frac{st}{u} \, \mathcal{T}_{\textrm{ad}}^{1243} \, I_4^6(s,u) \nonumber\\&&
 - 2 \left(  \frac{s^2}{t} \, \mathcal{T}_{\textrm{ad}}^{1234}
            +\frac{s^2}{u} \, \mathcal{T}_{\textrm{ad}}^{1243}
             \right) \, I_3^n(s) \nonumber\\&&
 -  2 \left(              s \, \mathcal{T}_{\textrm{ad}}^{1234}
             +\frac{s^2}{u} \, \mathcal{T}_{\textrm{ad}}^{1324} 
	     \right) \, I_3^n(t) 
 -  2 \left(              s \, \mathcal{T}_{\textrm{ad}}^{1243}
             +\frac{s^2}{t} \, \mathcal{T}_{\textrm{ad}}^{1324} 
	      \right) \, I_3^n(u) \nonumber\\&&
+   \left(-\frac{11\, s}{3\,t} \, \mathcal{T}_{\textrm{ad}}^{1234} 
          +\Bigl( \frac{11\, s}{3\,t} + \frac{t-u}{s}\Bigr) \, \mathcal{T}_{\textrm{ad}}^{1324} 
	  \right) \, I_2^n(t) \nonumber\\&&
+   \left(-\frac{11\, s}{3\,u} \, \mathcal{T}_{\textrm{ad}}^{1243}
          +\Bigl( \frac{11\, s}{3\,u} + \frac{u-t}{s}\Bigr) \, \mathcal{T}_{\textrm{ad}}^{1324}  
	   \right) \, I_2^n(u) \nonumber\\&&
- \left( \frac{s}{9\,t} \, \mathcal{T}_{\textrm{ad}}^{1234}	
        +\frac{s}{9\,u} \, \mathcal{T}_{\textrm{ad}}^{1243} 
	+\Bigl(\frac{s^2}{9\,tu} +1 \Bigr) \, \mathcal{T}_{\textrm{ad}}^{1324}
	   \right)  \;.
\eea	
In our definition we call the last line 
the rational part of the amplitude. Note that other polynomial contributions emerge 
after expanding the scalar integrals in $\epsilon$. 
We have checked that our result  is --- up to trivial factors stemming from  
different conventions --- identical 
to the amplitude representations provided in \cite{Bern:1991aq,Schubert:2001he}.

If we compute the same amplitude using off-shell values for the
external momenta, i.e. $k_1^2=k_2^2=k_3^2=k_4^2=m^2$, some scalar functions
change to their off-shell counterparts, e.g.
$I_3^n(s)\rightarrow I_3^n(s,m^2,m^2)$,
$I_4^{n+2}(u,t)\rightarrow I_4^{n+2}(u,t,m^2,m^2,m^2,m^2)$,
which are all IR finite now. 
The coefficients of the IR regulated basis integrals and the
constant part are up to terms of 
order $\mathcal{O}(m^2)$ identical to the coefficients of the original basis integrals.
The only difference stems from the no-scale 2-point functions which 
are replaced by their off-shell counterpart, $I_2^n(0)\rightarrow I_2^n(m^2)$.
This leads to an additional term 
which is proportional to the leading order amplitude:
\bea\label{off-shell-term}
 \left( \frac{32}{3} + \frac{4}{9} \epsilon +  \mathcal{O}(m^2)\right) \; I_2^n(m^2) \; \hat{\mathcal{A}}^{++--}_{\textrm{LO}} \;.
\eea
This additional term vanishes in the on-shell limit, as $I_2^n(0)=0$ in dimensional 
regularisation. When taking the on-shell limit it has to be put to zero {\em before} 
expanding in $\epsilon$, i.e. before the pole/rational part is extracted, otherwise 
the limit does not exist. 
This result defines the IR regulated version of the 
4-gluon amplitude. The IR-limit 
$\lim_{\textrm{IR}}\,(\Gamma_{\textrm{IR regulated}})=\Gamma$ is smooth as long as
scalar integrals are not expanded in $\epsilon$. For 2-point functions 
with  $\lim_{\textrm{IR}}\,I_2^n=0$ the rational/pole parts must not be isolated from 
the  result before the limit is taken. In practical calculations it is very easy to take
care of these terms separately. This reasoning shows that the rational part of an amplitude as defined above 
is not affected by IR divergences.

\subsection{The $\gamma\gamma ggg \to 0$ amplitude}

Following \cite{Binoth:2003xk} the helicity amplitudes
can be written as
\bea
\mathcal{M}^{\lambda_1\lambda_2\lambda_3\lambda_4\lambda_5} = \frac{Q_q^2 g_s^3}{i\pi^2} f^{c_3c_4c_5} 
\mathcal{A}^{\lambda_1\lambda_2\lambda_3\lambda_4\lambda_5}\; .
\eea
Due to Furry's theorem only one colour structure $\sim f^{c_3c_4c_5}$ exists.
The diagrammatic structure implies that
this amplitude is not cut constructible, as rank four and five 5-point functions
and rank three and four 4-point functions are present. 
Six independent helicity amplitudes exist. For three of them the cut-constructible
part is identically zero, they are given by the rational part only:
\bea
\mathcal{R}[\mathcal{A}^{+++++}] = \mathcal{A}^{+++++} = 
-\frac{ \textrm{tr}({\cal F}_1^+{\cal F}_2^+)\textrm{tr}({\cal F}_3^+{\cal F}_4^+{\cal F}_5^+) }{2\, s_{34}s_{45}s_{35}}\;.
\eea
Here ${\cal F}_j^{\mu\nu}=p_j^\mu\varepsilon_j^\nu-p_j^\nu\varepsilon_j^\mu$ is the abelian part of the gluon field strength tensor.
\begin{eqnarray}
{\cal A}^{-++++} = \frac{\textrm{tr}({\cal F}_2^+{\cal F}_3^+)\textrm{tr}({\cal F}_4^+{\cal F}_5^+) }{s_{23}^2 s_{45}^2}
\Bigl[ C^{-++++}\;  p_2\cdot {\cal F}_1^- \cdot p_4 - ( 4 \leftrightarrow 5 ) \Bigr]
\end{eqnarray} with the coefficient 
\begin{equation}
 C^{-++++} = -\frac{s_{15}s_{12}}{s_{24}s_{35}} 
                        -\frac{s_{15}}{s_{35}} + \frac{s_{23}}{s_{24}} - \frac{s_{15}}{s_{34}}\;.
\end{equation}
Further
\begin{eqnarray}
{\cal A}^{++++-} = \frac{\textrm{tr}({\cal F}_1^+{\cal F}_2^+)\textrm{tr}({\cal F}_3^+{\cal F}_4^+)}{s_{12}^2 s_{34}^2} 
\Bigl[ C^{++++-}  \; p_1\cdot {\cal F}_5^- \cdot p_3 - ( 3 \leftrightarrow 4 ) \Bigr]
\end{eqnarray} 
with the coefficient 
\begin{eqnarray}
C^{++++-} &=& -\frac{s_{45}s_{13}s_{14}}{s_{35}s_{15}s_{24}} 
-\frac{s_{13}s_{45}}{s_{15}s_{35}}
+\frac{s_{45}^{2}}{s_{15}s_{24}}
-\frac{s_{12}^{2}+s_{45}^{2}-s_{12}s_{45}}{s_{35}s_{15}}
+\frac{s_{13}s_{15}}{s_{23}s_{45}}
+\frac{s_{13}-s_{34}}{s_{23}} \nonumber\\ &&
-\frac{s_{34}s_{45}}{s_{23}s_{15}}
+\frac{s_{15}-s_{25}}{s_{45}}
-\frac{s_{23}+s_{35}}{s_{13}}
-\frac{s_{23}s_{25}}{s_{13}s_{45}}
+\frac{s_{34}+s_{12}}{s_{15}}\;.
\end{eqnarray}
The results for the helicity amplitudes
${\cal A}^{--+++}$, ${\cal A}^{+++--}$ and ${\cal A}^{-+++-}$
contain also contributions from the cuts. The full result for these
amplitudes can be found in \cite{Binoth:2003xk}. We have verified
that we get the same result for the rational terms, called 
${\cal A}^{--+++}_1$, ${\cal A}^{+++--}_1$ and ${\cal A}^{-+++-}_1$ there,
using our algebraic implementation of the rational polynomials
defined by our formalism.

\subsection{The 6-photon amplitude}

Due to Bose symmetry and parity invariance, only four independent
helicity amplitudes have to be evaluated, out of which two, the ``all plus" and the 
``one minus" amplitudes identically vanish~\cite{Mahlon:1993fe}.
We evaluated the 6-photon amplitudes along the lines of the 4-photon case, 
with the difference that after taking the trace, all
reducible scalar products in the numerator were cancelled directly.
We have verified that the rational parts of $\mathcal{M}^{++++++}$
and $\mathcal{M}^{+++++-}$ evaluate to zero using our formalism.
For the non-vanishing amplitudes, 
we find 
\bea
&&\mathcal{R}[ \mathcal{M}^{++++--} ] = 0 \label{4p2m}\\
&&\mathcal{R}[ \mathcal{M}^{+++---} ] = 0 \label{3p3m}\;,
\eea
where eq.~(\ref{4p2m}) is already known from the analytic result~\cite{Mahlon:1994dc}.
The results have been obtained by two independent calculations, 
one  based on the approach outlined in section 2, 
the other one  based on IR divergent form factors (see appendix \ref{sec:IR}).
We observe that for the  evaluation of the rational parts,  
form factors for at most rank 4 four-point functions were needed.
Kinematically they are of the same complexity as the ones used 
for the recent evaluation of the rational parts of the 
six gluon amplitude \cite{Xiao:2006vt}.

\section{Conclusions}

We have presented a formalism to evaluate the rational polynomials of arbitrary
one-loop $N$-point amplitudes. It is based on a tensor form factor representation 
derived in \cite{Binoth:2005ff}. The definition of rational
parts of these form factors induces a definition of the rational part
of the amplitude. 

To disentangle contributions form UV and IR poles, 
we  define first the rational polynomials with respect to 
IR-regulated amplitudes.
We obtain compact expressions for 
the rational and pole terms of
the tensor form factors which allow for an on-shell limit afterwards.
In this way it is clear that the  polynomial part of an amplitude
origins only from the UV behaviour of the amplitude.
This procedure defines rational polynomials of general 
one-loop amplitudes corresponding to the definition advocated in 
the literature~\cite{BDDK,Xiao:2006vr}.
In addition, we give all the formulae to work with
form factors which contain IR divergences in the appendix. 
For IR finite amplitudes both approaches obviously lead to the same
result.

Both approaches were implemented in algebraic manipulation programs
to allow for the fully automated evaluation of the rational parts
of one-loop amplitudes starting from Feynman diagrams.
The formalism has been applied to the evaluation  of the amplitudes for 
$gg\to H$, $\gamma\gamma \to \gamma\gamma$ , $gg \to gg$, $\gamma\gamma \to ggg$ and 
$\gamma\gamma \to \gamma\gamma\gamma\gamma$.
For the first four examples
we recover the well-known results. For the six-photon amplitude we have
proven by explicit analytical computation that the rational terms
of all Feynman diagrams add up to zero for all helicity configurations. 

Our implementation of this formalism is designed for arbitrary 
processes with up to six external legs, including massive particles, 
and makes it possible to obtain rational terms for
phenomenologically relevant partonic  amplitudes in an automated way. Note that 
numerical instabilities are typically mild in terms which are free of
logarithms and dilogarithms. The method is thus a  complement  to
the unitarity based techniques, which lead in general to
compact representations for coefficients of non-polynomial terms.
Combining  both methods thus might be a very fruitful starting
point for a highly effective  method to evaluate complex multi-leg 
one-loop amplitudes.

\section*{Acknowledgements}

First of all, we would like to thank Pierpaolo Mastrolia, because 
it is thanks to many interesting discussions with him that we  
focused particularly on this subject.  
T.B. and G.H. are grateful to the LAPTH Annecy for its hospitality 
while part of this work has been completed.
The work of T.B. was supported by the Deutsche Forschungsgemeinschaft (DFG)
under contract number BI 1050/1 and the Scottish Universities Physics Alliance (SUPA).
G.H. was supported by the Swiss National Science
Foundation (SNF) through grant no.\ 200021-101874.

\begin{appendix}
\renewcommand{\theequation}{\Alph{section}.\arabic{equation}}
\setcounter{equation}{0}

\section{Rational parts of IR divergent integrals}\label{sec:IR}

As an alternative to the approach outlined above, where IR regulated
amplitudes were considered, one can 
also define form factors for the rational parts in the presence of infrared poles. 
In section 2, we first apply the operator which extracts the rational parts 
and then take the on-shell limits of the form factors $F$:
\be
\lim_{\textrm{IR}}\left( {\cal P} + {\cal R}\right) \,\Big[ F(\textrm{IR-regulated}) \Big]\;.
\label{offshell}
\ee
Another possibility is to apply the operators ${\cal P}$ and ${\cal R}$  directly 
to the potentially IR-divergent form factors, i.e. do the operation 
\be
\left( {\cal P} + {\cal R}\right) \,\Big[ F(\textrm{IR-divergent})\Big]\;.
\label{onshell}
\ee


In an infrared finite amplitude, 
of course  all finite terms 
coming from the expansion of $\eps$-dependent terms  combined 
with $1/\eps_{\rm{IR}}$ poles finally 
have to cancel, such that the 
{\it remaining} finite polynomial parts
are  identical to the ones obtained by procedure (\ref{offshell}), 
after summing over all contributions. 
In an infrared divergent amplitude, the finite remainders are related 
to the choice of the factorisation scheme.

In order to be able to define the polynomial part of divergent 
amplitudes, it is necessary to single out the contributions which come from
the expansion of the poles. 
The results of operation (\ref{onshell}) for the divergent three-point functions with Feynman 
parameters in the numerator, shown in Tables 
\ref{table1} to \ref{table3} below, are given in a form which allows to 
isolate these contributions immediately.  
 
The definition of the rational parts of  
expressions containing $1/\eps^2$ and $1/\eps$ poles is of course linked to the overall 
$\eps$-dependent factors which have been extracted. 
We single out the pole contributions in terms of 
${\cal U}[I_3^n(0,0,X)]$ and ${\cal U}[I_2^n]$, i.e. 
all double poles have been absorbed 
into the scalar three-point function with two light-like legs, 
depending only on the invariant $X$,  and all 
single poles have been absorbed into $I_2^{n}$, where 
\bea
I_2^n\equiv I_2^n(X)&=&\frac{\tilde{r}_\Gamma}{\eps}(-X)^{-\eps}\\
I_3^n\equiv I_3^n(0,0,X)&=&\frac{\tilde{r}_\Gamma}{\eps^2}\frac{(1-2\eps)}{X}\,(-X)^{-\eps}\\
\tilde{r}_\Gamma & = & \frac{\Gamma(1 +\eps) \,
\Gamma^2(1-\eps)}{\Gamma(2-2 \,\eps)}=\frac{r_\Gamma}{1-2\eps}\;.
\eea
Extracting an overall factor $\tilde{r}_\Gamma$ instead of 
${r}_\Gamma$ from {\it all} integrals, we have 
$\mathcal{R}[I_2^n] = 0$, which is more convenient 
for our purposes than extracting an overall factor 
$\tilde{r}_\Gamma$, which would imply $\mathcal{R}[I_2^n] = 2$. 

For three-point functions with one non-zero invariant $X$, 
we labelled the internal propagators in such a way that 
$\cals_{1 3}=X$ and $\cals_{1 2}=\cals_{2 3}=0$, where $\cals$ is defined in 
eq.\,(\ref{fofa}). 
For three-point functions with two non-zero invariants $X$ and $Y$, 
we set 
$\cals_{2 3} = X$ and $\cals_{1 3} = Y$. 
Thus the integrals $I^n_3(i,j,\ldots;0,0,X)$
are symmetric under exchange of 
$1\leftrightarrow 3$ and the $I^n_3(i,j,\ldots;0,X,Y)$ 
are symmetric under 
simultaneous exchange of 
$1\leftrightarrow 2$ and $X\leftrightarrow Y$. 

In the tables, the results for the  
divergent three-point functions with Feynman parameters in the numerator 
are split up in the following way: 
\bea
\mathcal{U}[
I_3^{n}(\{j_l\})]&=&
d_{3}\,{\cal U}[I_3^{n}(0,0,X)]+
d_{2}\,{\cal U}[I_2^{n}]+{\cal V}(\{j_l\})+W(\{j_l\})\;, 
\label{cd}
\eea
where $d_{3}$ is equal to one if the integral has a double pole, and zero 
otherwise.  
The functions 
${\cal V}(\{j_l\})$ and $W(\{j_l\})$ denote the remaining finite part, 
where ${\cal V}(\{j_l\})$ is the part which is equal to the limit 
of the corresponding expression for the off-shell integral 
(given in eqs.~(\ref{b32}),(\ref{AB3}))
when one or two invariants go to zero, and
$W(\{j_l\})$ is the finite remainder, i.e. the {\it difference} to the 
on-shell limit of the corresponding off-shell integral, after having singled 
out the pole contributions.  
To give an example, we have 
\bea
&&{\cal U}[I_3^{n}(2,2,2;0,0,s_1)]={\cal U}[I_3^{n}(0,0,s_1)]+
\frac{11}{3s_1}\,{\cal U}[I_2^{n}]+\frac{19}{9s_1}\nn\\
&&\lim_{s_2,s_3\to 0}\,{\cal U}[I_3^{n}(2,2,2;s_3,s_2,s_1)]=
-\lim_{s_2,s_3\to 0}\,{\cal U}[A_{222}^{3,3}]=\frac{2}{s_1}=
{\cal V}(2,2,2;s_1)\nn\\
&\Rightarrow& W(2,2,2;s_1)=\frac{19}{9s_1}-{\cal V}(2,2,2;s_1)=\frac{1}{9s_1}\;.
\eea
The fact that $W$ is not always zero of course does not mean that 
the results obtained by procedures (\ref{offshell}) and (\ref{onshell}) 
have to be different. 
The coefficients of the corresponding integrals 
as well as the number of non-zero two-point functions are different 
in the two approaches, such that after summation over all contributions 
making up a finite amplitude, the results will be the same.
This has been checked explicitly by calculating the 4-photon and the 6-photon 
amplitudes in both ways.


\TABLE{
\begin{tabular}{|l|c|c|c|}
\hline
&&&\\
&${\cal U}[I]$&${\cal V}$&$W$\\
&&&\\
\hline 
\hline
&&&\\
$I^n_3(0,0,X)$ & ${\cal U}[I_3^n]$&$0$&0\\
&&&\\
\hline
&&&\\
$I^n_3(1;0,0,X)$ &$-\frac{1}{X}\,{\cal U}[I_2^n]$&0&0\\[2mm]
$I^n_3(2;0,0,X)$ &${\cal U}[I_3^n]+\frac{2}{X}\,{\cal U}[I_2^n]$&0&0\\[2mm]
$I^n_3(3;0,0,X)$ &$-\frac{1}{X}\,{\cal U}[I_2^n]$&0&0\\
&&&\\
\hline
&&&\\
$I^n_3(1,1;0,0,X)$ & $-\frac{1}{2X}\,{\cal U}[I_2^n]$&0&0\\[2mm]
$I^n_3(2,2;0,0,X)$ & ${\cal U}[I_3^n]+\frac{3}{X}\,{\cal U}[I_2^n]+\frac{1}{X}$&$\frac{1}{X}$&0\\[2mm]
$I^n_3(3,3;0,0,X)$ & $-\frac{1}{2X}\,{\cal U}[I_2^n]$&0&0\\[2mm]
$I^n_3(1,2;0,0,X)$ & $-\frac{1}{2X}\,{\cal U}[I_2^n]-\frac{1}{2X}$&$-\frac{1}{2X}$&0\\[2mm]
$I^n_3(1,3;0,0,X)$ & $\frac{1}{2X}$ & $\frac{1}{2X}$  &0\\[2mm]
$I^n_3(2,3;0,0,X)$ & $-\frac{1}{2X}\,{\cal U}[I_2^n]-\frac{1}{2X}$&$-\frac{1}{2X}$&0\\
&&&\\
\hline
&&&\\
$I^n_3(1,1,1;0,0,X)$ & $-\frac{1}{3X}\,{\cal U}[I_2^n]-\frac{1}{18X}$& 0&$-\frac{1}{18X}$\\[2mm]
$I^n_3(2,2,2;0,0,X)$ & ${\cal U}[I_3^n]+\frac{11}{3X}\,{\cal U}[I_2^n]+
\frac{19}{9X}$& $\frac{2}{X}$&$\frac{1}{9X}$\\[2mm]
$I^n_3(3,3,3;0,0,X)$ &$-\frac{1}{3X}\,{\cal U}[I_2^n]-\frac{1}{18X}$& 0&
$-\frac{1}{18X}$\\[2mm]
$I^n_3(1,1,2;0,0,X)$ & $-\frac{1}{6X}\,{\cal U}[I_2^n]-\frac{1}{9X}$&
$-\frac{1}{6X}$ &$\frac{1}{18X}$\\[2mm]
$I^n_3(1,2,2;0,0,X)$ & $-\frac{1}{3X}\,{\cal U}[I_2^n]-\frac{5}{9X}$&
$-\frac{1}{2X}$ &$-\frac{1}{18X}$\\[2mm]
$I^n_3(1,1,3;0,0,X)$ &$\frac{1}{6X}$&$\frac{1}{6X}$&0\\[2mm]
$I^n_3(2,2,3;0,0,X)$ & $-\frac{1}{3X}\,{\cal U}[I_2^n]-\frac{5}{9X}$&
$-\frac{1}{2X}$ &$-\frac{1}{18X}$\\[2mm]
$I^n_3(1,3,3;0,0,X)$ &$\frac{1}{6X}$&$\frac{1}{6X}$&0\\[2mm]
$I^n_3(2,3,3;0,0,X)$ & $-\frac{1}{6X}\,{\cal U}[I_2^n]-\frac{1}{9X}$&
$-\frac{1}{6X}$ &$\frac{1}{18X}$\\[2mm]
$I^n_3(1,2,3;0,0,X)$&$\frac{1}{6X}$&$\frac{1}{6X}$&0 \\
&&&\\
\hline
\end{tabular}
\caption{Rational and pole parts of three-point functions with two light-like 
legs and up to three Feynman parameters in the numerator. ${\cal U}$ 
is the operator extracting the pole and rational parts of the 
integral. The $1/\eps$ and $1/\eps^2$ poles have been absorbed in the terms 
proportional to ${\cal U}[I_2^n]$ and ${\cal U}[I_3^n]$, respectively. 
${\cal V}$ denotes the value 
obtained from the corresponding expression for the off-shell integral, 
in the limit where two legs go on-shell. $W$ is the difference 
${\cal U}[I]-{\cal V}$, where the pole terms have been set to zero.}\label{table1}
}

\TABLE{
\begin{tabular}{|l|c|c|c|}
\hline
&&&\\
&${\cal U}[I]$&${\cal V}$&$W$\\
&&&\\
\hline 
\hline &&&\\
$I^n_3(0,X,Y)$ & 0 & 0&0\\
&&&\\
\hline &&&\\
$I^n_3(1;0,X,Y)$ & $\frac{1}{X-Y}\,{\cal U}[I_2^n]$ & 0&0\\[2mm]
$I^n_3(2;0,X,Y)$ & $\frac{1}{Y-X} \,{\cal U}[I_2^n]$ & 0&0\\[2mm]
$I^n_3(3;0,X,Y)$ & 0 & 0&0\\
&&&\\
\hline &&&\\
$I^n_3(1,1;0,X,Y)$ & $\frac{3X-Y}{2(X-Y)^2}\,{\cal U}[I_2^n]+\frac{X}{(X-Y)^2}$&$\frac{X}{(X-Y)^2}$&0\\[2mm]
$I^n_3(2,2;0,X,Y)$ & $\frac{3Y-X}{2(X-Y)^2}\,{\cal U}[I_2^n]+\frac{Y}{(X-Y)^2}$&$\frac{Y}{(X-Y)^2}$&0\\[2mm]
$I^n_3(3,3;0,X,Y)$ &  0 & 0&0\\[2mm]
$I^n_3(1,2;0,X,Y)$ & $-\frac{X+Y}{2(X-Y)^2}\,\left({\cal U}[I_2^n]+1\right)$&$-\frac{X+Y}{2(X-Y)^2}$&0\\[2mm]
$I^n_3(1,3;0,X,Y)$ & $-\frac{1}{2(X-Y)}$& $-\frac{1}{2(X-Y)}$&0\\[2mm]
$I^n_3(2,3;0,X,Y)$ & $-\frac{1}{2(Y-X)}$& $-\frac{1}{2(Y-X)}$&0\\
&&&\\
\hline &&&\\
$I^n_3(1,1,1;0,X,Y)$ & $\frac{11X^2-7XY+2Y^2}{6(X-Y)^3}\,{\cal U}[I_2^n]+\frac{37X^2-8XY+Y^2}{18(X-Y)^3}$& $\frac{X(6X-Y)}{3(X-Y)^3}$&$\frac{1}{18(X-Y)}$\\[2mm]
$I^n_3(2,2,2;0,X,Y)$ &$\frac{11Y^2-7XY+2X^2}{6(Y-X)^3}\,{\cal U}[I_2^n]+\frac{37Y^2-8XY+X^2}{18(Y-X)^3}$&$\frac{Y(6Y-X)}{3(Y-X)^3}$&$\frac{1}{18(Y-X)}$\\[2mm]
$I^n_3(3,3,3;0,X,Y)$ &0&0&0\\[2mm]
$I^n_3(1,1,2;0,X,Y)$ &$
\frac{-2X^2-5XY+Y^2}{6(X-Y)^3}\,{\cal U}[I_2^n]+\frac{-5X^2-11XY+Y^2}{9(X-Y)^3}$&$\frac{-3X^2-8XY+Y^2}{6(X-Y)^3}$&$-\frac{1}{18(X-Y)}$\\[2mm]
$I^n_3(1,2,2;0,X,Y)$ &$\frac{-2Y^2-5XY+X^2}{6(Y-X)^3}\,{\cal U}[I_2^n]+\frac{-5Y^2-11XY+X^2}{9(Y-X)^3}$&$\frac{-3Y^2-8XY+X^2}{6(Y-X)^3}$ &$-\frac{1}{18(Y-X)}$\\[2mm]
$I^n_3(1,1,3;0,X,Y)$ &$\frac{-3X+Y}{6(X-Y)^2}$&$\frac{-3X+Y}{6(X-Y)^2}$&0\\[2mm]
$I^n_3(2,2,3;0,X,Y)$ &$\frac{-3Y+X}{6(X-Y)^2}$&$\frac{-3Y+X}{6(X-Y)^2}$&0\\[2mm]
$I^n_3(1,3,3;0,X,Y)$ &$-\frac{1}{6(X-Y)}$&$-\frac{1}{6(X-Y)}$&0\\[2mm] 
$I^n_3(2,3,3;0,X,Y)$ &$-\frac{1}{6(Y-X)}$&$-\frac{1}{6(Y-X)}$&0\\[2mm]
$I^n_3(1,2,3;0,X,Y)$ &$\frac{(X+Y)}{6(X-Y)^2}$&$\frac{(X+Y)}{6(X-Y)^2}$&0\\
&&&\\
\hline
\end{tabular}
\caption{Rational and pole parts of three-point functions with one light-like 
leg and up to three Feynman parameters in the numerator. ${\cal U}$ 
is the operator extracting the pole and rational parts of the 
integral.  ${\cal V}$ denotes the value 
obtained from the corresponding expression for the off-shell integral, 
in the limit where one leg goes on-shell. $W$ is the difference 
${\cal U}[I]-{\cal V}$ where the pole terms, absorbed into  ${\cal U}[I_2^n]$, 
have been set to zero.}\label{table2}
}

\TABLE{
\begin{tabular}{|l|c|c|c|}
\hline 
&&&\\
&${\cal U}[I]$&${\cal V}$&$W$\\
&&&\\
\hline 
\hline &&&\\
$I_3^{n+2}(0,0,X)$&$-\frac{1}{2}{\cal U}[I_2^n]-\frac{1}{2}$&
$-\frac{1}{2}{\cal U}[I_2^n]-\frac{1}{2}$&0\\
&&&\\
\hline &&&\\
$I_3^{n+2}(1;0,0,X)$&$-\frac{1}{6}{\cal U}[I_2^n]-\frac{1}{9}$&$-\frac{1}{6}{\cal U}[I_2^n]-\frac{1}{9}$&0\\[2mm]
$I_3^{n+2}(2;0,0,X)$&$-\frac{1}{6}{\cal U}[I_2^n]-\frac{5}{18}$&$-\frac{1}{6}{\cal U}[I_2^n]-\frac{5}{18}$&0\\[2mm]
$I_3^{n+2}(3;0,0,X)$&$-\frac{1}{6}{\cal U}[I_2^n]-\frac{1}{9}$&$-\frac{1}{6}{\cal U}[I_2^n]-\frac{1}{9}$&0\\
&&&\\
\hline 
\hline &&&\\
$I_3^{n+2}(0,X,Y)$&$-\frac{1}{2}{\cal U}[I_2^n]-\frac{1}{2}$&
$-\frac{1}{2}{\cal U}[I_2^n]-\frac{1}{2}$&0\\
&&&\\
\hline &&&\\
$I_3^{n+2}(1;0,X,Y)$&$-\frac{1}{6}\,{\cal U}[I_2^n]+\frac{-5X+2Y}{18(X-Y)}$&$-\frac{1}{6}\,{\cal U}[I_2^n]+\frac{-5X+2Y}{18(X-Y)}$&0\\[2mm]
$I_3^{n+2}(2;0,X,Y)$&$-\frac{1}{6}\,{\cal U}[I_2^n]+\frac{-5Y+2X}{18(Y-X)}$&$-\frac{1}{6}\,{\cal U}[I_2^n]+\frac{-5X+2Y}{18(X-Y)}$&0\\[2mm]
$I_3^{n+2}(3;0,X,Y)$&$-\frac{1}{6}\,{\cal U}[I_2^n]-\frac{1}{9}\qquad\quad$&$-\frac{1}{6}\,{\cal U}[I_2^n]-\frac{1}{9}\qquad\quad$&0\\
&&&\\
\hline
\end{tabular}
\caption{Rational and pole parts of $(n+2)$-dimensional three-point functions with one or two  
light-like 
legs and up to one Feynman parameter in the numerator. ${\cal U}$ 
is the operator extracting the pole and rational parts of the 
integral. 
Note that  the poles in $I_3^{n+2}$ are of ultraviolet 
nature, but we do not distinguish the nature of the poles 
denoted by $\mathcal{P}[I_2^n]$.
${\cal V}$ denotes the value 
obtained from the corresponding expression for the off-shell integral, 
in the limit where one or two legs go on-shell. $W$ is the difference 
${\cal U}[I]-{\cal V}$, where the pole terms, 
absorbed into  ${\cal U}[I_2^n]$, are set to zero.}\label{table3}
}

\end{appendix}


\clearpage

\begin{thebibliography}{99}

\bibitem{houches05}
C.~Buttar et al., 
{\it ``Les Houches workshop on Physics at TeV Colliders 2005, Standard Model, QCD, EW, and
  Higgs working group: Summary report''},
  hep-ph/0604120.


\bibitem{Soper:1999xk}
  D.~E.~Soper,
  Phys.\ Rev.\ D {\bf 62} (2000) 014009 [hep-ph/9910292];\\
  M.~Kr\"amer and D.~E.~Soper,
  Phys.\ Rev.\ D {\bf 66} (2002) 054017
  [hep-ph/0204113].


\bibitem{Ferroglia:2002mz}
  A.~Ferroglia, M.~Passera, G.~Passarino and S.~Uccirati,
  Nucl.\ Phys.\ B {\bf 650} (2003) 162
  [hep-ph/0209219].

\bibitem{Nagy:2003qn}
  Z.~Nagy and D.~E.~Soper,
  JHEP {\bf 0309} (2003) 055
  [hep-ph/0308127].

\bibitem{Kurihara:2005ja}
  Y.~Kurihara and T.~Kaneko,
  Comput.\ Phys.\ Commun.\  {\bf 174} (2006) 530
  [hep-ph/0503003].

\bibitem{Anastasiou:2005cb}
  C.~Anastasiou and A.~Daleo,
  hep-ph/0511176.
  
\bibitem{Czakon:2005rk}
  M.~Czakon,
  Comput.\ Phys.\ Commun.\  {\bf 175} (2006) 559
  [hep-ph/0511200].


\bibitem{Binoth:2002xh}
  T.~Binoth, G.~Heinrich and N.~Kauer,
  Nucl.\ Phys.\ B {\bf 654} (2003) 277
  [hep-ph/0210023].

\bibitem{Grace}
G.~B\'elanger, F.~Boudjema, J.~Fujimoto, T.~Ishikawa, T.~Kaneko, 
K.~Kato and Y.~Shimizu,
Phys.\ Rept.\  {\bf 430}, 117 (2006)
[hep-ph/0308080].

\bibitem{delAguila:2004nf}
  F.~del Aguila and R.~Pittau,
  JHEP {\bf 0407} (2004) 017
  [hep-ph/0404120].

\bibitem{vanHameren:2005ed}
  A.~van Hameren, J.~Vollinga and S.~Weinzierl,
  Eur.\ Phys.\ J.\ C {\bf 41} (2005) 361
  [hep-ph/0502165].

\bibitem{Binoth:2005ff}
  T.~Binoth, J.~P.~Guillet, G.~Heinrich, E.~Pilon and C.~Schubert,
  JHEP {\bf 0510} (2005) 015
  [hep-ph/0504267].

\bibitem{Ellis:2005zh}
  R.~K.~Ellis, W.~T.~Giele and G.~Zanderighi,
  Phys.\ Rev.\ D {\bf 73} (2006) 014027
  [hep-ph/0508308].

\bibitem{Denner:2005nn}
  A.~Denner and S.~Dittmaier,
  Nucl.\ Phys.\ B {\bf 734} (2006) 62
  [hep-ph/0509141].
  
\bibitem{Binoth:2006rc}
  T.~Binoth, M.~Ciccolini and G.~Heinrich,
  hep-ph/0601254.
  
\bibitem{Binoth:2006mf}
  T.~Binoth, A.~Guffanti, J.~P.~Guillet, S.~Karg, N.~Kauer and T.~Reiter,
  hep-ph/0606318.
  

\bibitem{witten}
E.~Witten,
Commun.\ Math.\ Phys.\ {\bf 252}, 189 (2004)
[hep-th/0312171].

\bibitem{BST}
A.~Brandhuber, B.~Spence and G.~Travaglini,
Nucl.\ Phys.\ B {\bf 706}, 150 (2005)
[hep-th/0407214].

\bibitem{BCF7}
R.~Britto, F.~Cachazo and B.~Feng,
Phys.\ Rev.\ D {\bf 71}, 025012 (2005)
[hep-th/0410179].

\bibitem{BBST}
J.~Bedford, A.~Brandhuber, B.~Spence and G.~Travaglini,
Nucl.\ Phys.\ B {\bf 706}, 100 (2005)
[hep-th/0410280];\\
J.~Bedford, A.~Brandhuber, B.~Spence and G.~Travaglini,
Nucl.\ Phys.\ B {\bf 712}, 59 (2005)
[hep-th/0412108].
%
\bibitem{altogether}
S.~J.~Bidder, N.~E.~J.~Bjerrum-Bohr, L.~J.~Dixon and D.~C.~Dunbar,
Phys.\ Lett.\ B {\bf 606}, 189 (2005)
[hep-th/0410296];\\
%
S.~J.~Bidder, N.~E.~J.~Bjerrum-Bohr, D.~C.~Dunbar and W.~B.~Perkins,
Phys.\ Lett.\ B {\bf 608}, 151 (2005)
[hep-th/0412023];\\
%
S.~J.~Bidder, N.~E.~J.~Bjerrum-Bohr, D.~C.~Dunbar and W.~B.~Perkins,
Phys.\ Lett.\ B {\bf 612}, 75 (2005)
[hep-th/0502028].

\bibitem{BBCF}
R.~Britto, E.~Buchbinder, F.~Cachazo and B.~Feng,
Phys.\ Rev.\ D {\bf 72}, 065012 (2005)
[hep-ph/0503132].

\bibitem{Bern:2005hh}
  Z.~Bern, N.~E.~J.~Bjerrum-Bohr, D.~C.~Dunbar and H.~Ita,
  JHEP {\bf 0511} (2005) 027
  [hep-ph/0507019].
%
\bibitem{Britto:2006sj}
R.~Britto, B.~Feng and P.~Mastrolia,
Phys.\ Rev.\ D {\bf 73}, 105004 (2006)
[hep-ph/0602178].

\bibitem{Bern:1994zx}
  Z.~Bern, L.~J.~Dixon, D.~C.~Dunbar and D.~A.~Kosower,
  Nucl.\ Phys.\ B {\bf 425} (1994) 217
  [hep-ph/9403226].

\bibitem{BDDK}
  Z.~Bern, L.~J.~Dixon, D.~C.~Dunbar and D.~A.~Kosower,
  Nucl.\ Phys.\ B {\bf 435} (1995) 59
  [hep-ph/9409265].

\bibitem{moreUnitarity}
Z.\ Bern and A.\ G.\ Morgan,
Nucl.\ Phys.\ B {\bf 467}, 479 (1996)
[hep-ph/9511336];\\
%
Z.\ Bern, L.\ J.\ Dixon and D.\ A.\ Kosower,
Ann.\ Rev.\ Nucl.\ Part.\ Sci.\ {\bf 46}, 109 (1996)
[hep-ph/9602280].

\bibitem{NeqFourSevenPoint}
Z.~Bern, V.~Del Duca, L.~J.~Dixon and D.~A.~Kosower,
Phys.\ Rev.\ D {\bf 71}, 045006 (2005)
[hep-th/0410224].

\bibitem{BCFII}
R.~Britto, F.~Cachazo and B.~Feng,
Nucl.\ Phys.\ B {\bf 725}, 275 (2005)
[hep-th/0412103].

\bibitem{NeqFourNMHV}
Z.~Bern, L.~J.~Dixon and D.~A.~Kosower,
Phys.\ Rev.\ D {\bf 72}, 045014 (2005)
[hep-th/0412210].


\bibitem{Ellis:2006ss}
  R.~K.~Ellis, W.~T.~Giele and G.~Zanderighi,
  JHEP {\bf 0605} (2006) 027
  [hep-ph/0602185].
  
\bibitem{Xiao:2006vr}
  Z.~G.~Xiao, G.~Yang and C.~J.~Zhu,
  hep-ph/0607015.

\bibitem{Su:2006vs}
  X.~Su, Z.~G.~Xiao, G.~Yang and C.~J.~Zhu,
  hep-ph/0607016.

\bibitem{Xiao:2006vt}
  Z.~G.~Xiao, G.~Yang and C.~J.~Zhu,
  hep-ph/0607017.

\bibitem{Bern:1993kr}
Z.~Bern, L.~J.~Dixon and D.~A.~Kosower,
Nucl.\ Phys.\ B {\bf 412} (1994) 751
[hep-ph/9306240].
\bibitem{Bern:1992em}
Z.~Bern, L.~J.~Dixon and D.~A.~Kosower,
Phys.\ Lett.\ B {\bf 302} (1993) 299
[Erratum-ibid.\ B {\bf 318} (1993) 649]
[hep-ph/9212308].


\bibitem{Bern:2005hs}
Z.~Bern, L.~J.~Dixon and D.~A.~Kosower,
Phys.\ Rev.\ D {\bf 71}, 105013 (2005)
[hep-th/0501240].


\bibitem{Bern:2005ji}
Z.~Bern, L.~J.~Dixon and D.~A.~Kosower,
Phys.\ Rev.\ D {\bf 72}, 125003 (2005)
[hep-ph/0505055].

\bibitem{Bern:2005cq}
  Z.~Bern, L.~J.~Dixon and D.~A.~Kosower,
  Phys.\ Rev.\ D {\bf 73} (2006) 065013
  [hep-ph/0507005].


\bibitem{Berger:2006ci}
C.~F.~Berger, Z.~Bern, L.~J.~Dixon, D.~Forde and D.~A.~Kosower,
hep-ph/0604195.

\bibitem{Forde:2005hh}
  D.~Forde and D.~A.~Kosower,
  Phys.\ Rev.\ D {\bf 73} (2006) 061701
  [hep-ph/0509358].

\bibitem{Berger:2006vq}
C.~F.~Berger, Z.~Bern, L.~J.~Dixon, D.~Forde and D.~A.~Kosower,
hep-ph/0607014.
%
\bibitem{Berger:2006sh}
  C.~F.~Berger, V.~Del Duca and L.~J.~Dixon,
  hep-ph/0608180.

\bibitem{Ossola:2006us}
  G.~Ossola, C.~G.~Papadopoulos and R.~Pittau,
  hep-ph/0609007.

\bibitem{Badger:2006us}
  S.~D.~Badger and E.~W.~N.~Glover,
  hep-ph/0607139.


\bibitem{davy}
A.~I.~Davydychev,
Phys.\ Lett.\ B {\bf 263}, 107 (1991).

\bibitem{Binoth:1999sp}
  T.~Binoth, J.~P.~Guillet and G.~Heinrich,
  Nucl.\ Phys.\ B {\bf 572} (2000) 361 [hep-ph/9911342].

\bibitem{Giele:2004iy}
  W.~T.~Giele and E.~W.~N.~Glover,
  JHEP {\bf 0404} (2004) 029
  [hep-ph/0402152].
  
\bibitem{Kunszt:1993sd}
  Z.~Kunszt, A.~Signer and Z.~Trocsanyi,
  Nucl.\ Phys.\ B {\bf 411} (1994) 397
  [hep-ph/9305239].
\bibitem{Catani:1996pk}
  S.~Catani, M.~H.~Seymour and Z.~Trocsanyi,
  Phys.\ Rev.\ D {\bf 55} (1997) 6819
  [hep-ph/9610553].

\bibitem{Smith:2004ck}
  J.~Smith and W.~L.~van Neerven,
  Eur.\ Phys.\ J.\ C {\bf 40} (2005) 199
  [hep-ph/0411357].

\bibitem{Veltman:1988au}
  M.~J.~G.~Veltman,
  Nucl.\ Phys.\ B {\bf 319} (1989) 253.

  
\bibitem{xuetal} Z.~Xu, D.~Zhang, L.~Chang, Nucl.\ Phys.\ B {\bf 291} (1987) 392.
  
\bibitem{Gastmans:1990xh}
  R.~Gastmans and T.~T.~Wu,
  {\it ``The Ubiquitous Photon: Helicity Method For QED And QCD''},
 Oxford, UK: Clarendon (1990), 
 International series of monographs on physics, 80.  

\bibitem{Binoth:2002xg}
  T.~Binoth, E.~W.~N.~Glover, P.~Marquard and J.~J.~van der Bij,
  JHEP {\bf 0205} (2002) 060
  [hep-ph/0202266].

\bibitem{Binoth:2003xk}  
  T.~Binoth, J.~P.~Guillet and F.~Mahmoudi,
  JHEP {\bf 0402} (2004) 057 [hep-ph/0312334].

\bibitem{Bern:1991aq}
  Z.~Bern and D.~A.~Kosower,
  Nucl.\ Phys.\ B {\bf 379} (1992) 451.

\bibitem{Schubert:2001he}
  C.~Schubert,
  Phys.\ Rept.\  {\bf 355} (2001) 73
  [hep-th/0101036].

\bibitem{Mahlon:1993fe}
  G.~Mahlon,
  Phys.\ Rev.\ D {\bf 49} (1994) 2197
  [hep-ph/9311213].

\bibitem{Mahlon:1994dc}
  G.~Mahlon,
  {\it Talk given at 4th International Conference on Physics Beyond the Standard Model, 
  Lake Tahoe, CA, Dec 1994}, 
FERMILAB-Conf-94-421, hep-ph/9412350.



\end{thebibliography}
\end{document}